%% file: HoughS5_CQG_v3.tex
\documentclass[10pt]{iopart}

\usepackage{iopams}  

\usepackage{graphicx}
\usepackage{hyperref}
\usepackage{amssymb}
\usepackage{longtable}
\usepackage{rotating}
\usepackage{color}
\usepackage{epsfig}
\usepackage{epsf}
\usepackage{fancyhdr}

\def\be{\begin{equation}}
\def\ee{\end{equation}}

\def\beq{\begin{equation}}
\def\eeq{\end{equation}}
\def\bea{\begin{eqnarray}}
\def\eea{\end{eqnarray}}

\def\H{\textrm{\mbox{\tiny{H}}}}
\def\th{\textrm{\mbox{\tiny{th}}}}
\def\Tobs{T_{\textrm{\mbox{\tiny{obs}}}}}
\def\Tcoh{T_{\textrm{\mbox{\tiny{coh}}}}}

\def\iSubSupInd{i}
\def\srchTemplateInd{{}}

\newcommand{\ba}{\begin{eqnarray}}
\newcommand{\ea}{\end{eqnarray}}

\newcommand{\bml}{\begin{mathletters}}
\newcommand{\eml}{\end{mathletters}}

\def \S {{\cal S}}

\def\sci#1#2{#1\times10^{#2}}

\begin{document}

\title[Hough search for continuous gravitational waves in LIGO S5 data]
{Application of a Hough search for continuous gravitational waves on data from the 5th LIGO science run}


\input{authors_LSC_Feb2013_Virgo_Jul2013_iop.tex}
\ead{alicia.sintes@uib.es}

\begin{abstract}
We report on an all--sky search for periodic gravitational waves in the frequency range $\mathrm{50-1000\,Hz}$
with the first derivative of frequency in the range $-8.9 \times 10^{-10}$~Hz/s to zero 
in two years of data collected during LIGO's fifth science run. 
Our results employ a Hough transform technique, 
introducing a $\chi^2$ test and analysis of coincidences between the signal levels in years
1 and 2 of observations that offers a significant improvement in the product of strain sensitivity with compute cycles per data sample
compared to previously published searches.
Since our search yields no surviving candidates, we present results taking the form of frequency dependent, 95$\%$ 
confidence upper limits on the strain amplitude $h_0$. The most stringent upper limit from year 1 is $1.0\times 10^{-24}$
in the $\mathrm{158.00-158.25\,Hz}$ band. In year 2, the most stringent upper limit is $\mathrm{8.9\times10^{-25}}$ in the
$\mathrm{146.50-146.75\,Hz}$ band. 
This improved detection pipeline, which is computationally efficient by at least two orders of magnitude better than our flagship  Einstein$@$Home search, will be important for ``quick-look'' searches in the Advanced LIGO and Virgo detector era.

\end{abstract}




\section{Introduction}
\label{sec:introduction}


The focus of this article is the search for evidence of continuous gravitational waves, as might be radiated by nearby, rapidly spinning neutron stars, in  data from the Laser Interferometer Gravitational-wave Observatory (LIGO) \cite{Abbott:2007kv}.  The data used in this paper were produced during LIGO's fifth science run (S5) that started on November 4, 2005 and ended on October 1, 2007.  

Spinning neutron stars are promising sources of gravitational wave signals in the LIGO frequency band. These objects may generate continuous gravitational waves  through a variety of mechanisms including non-axisymmetric distortions of the neutron star, unstable oscillation modes in the fluid part of the star and free precession \cite{Bildsten:1998ey, Ushomirsky:2000ax, Cutler:2002nw, Melatos:2005ez, Owen:2005fn}. Independently of the specific mechanism, the emitted signal is a quasi-periodic wave whose frequency changes slowly during the observation time due to energy loss through gravitational wave emission, and possibly other mechanisms. At an Earth-based detector the signal exhibits amplitude and phase modulations due to the motion of the Earth with respect to the source.

A number of searches have been carried out previously in LIGO data \cite{Abbott:2005pu, Abbott:2006vg, Abbott:2007ce, :2007tda, Abbott:2008fx,  Abbott:2008uq, :2008rg, Collaboration:2009nc, Collaboration:2009rfa, Abadie:2011md, Abadie:2011wj, Aasi:2012fw}  including:  targeted searches in which precise pulsar ephemerides from radio, X-ray or $\gamma$-ray observations can be used in a coherent  integration over the full observation span; directed searches  in which the direction of the source is known precisely, but for which little or no frequency information is known; and all-sky searches in which there is no information about location or frequency.

All-sky searches for unknown neutron stars must cope with a very large parameter space volume. Optimal methods based on 
coherent integration over the full observation time are completely unfeasible since the template bank spacing decreases dramatically  with observation time, and even for a coherent time baseline of just  few days, a wide-frequency-band all-sky search is computationally extremely challenging. Therefore hierarchical approaches have been
 proposed~\cite{Brady:1998nj, Papa:2000wg, Krishnan:2004sv, Cutler:2005pn, Prix:2012yu, Shaltev:2013kqa} 
 which incorporate semi-coherent methods into the analysis.
 These techniques  
 are less sensitive for the same observation time but are computationally inexpensive. 
 The Hough transform~\cite{ Krishnan:2004sv, Abbott:2005pu, Sintes:2006uc, :2007tda,  Palomba:2005fp} 
 is an example of such a method and has been used in previous wide-parameter-space searches published by the LIGO and Virgo Collaborations. Moreover it has also  been used in the hierarchical approach 
 for Einstein$@$Home searches, as the incoherent method to combine the information from coherently analyzed 
 segments \cite{Collaboration:2009nc, Aasi:2012fw}. 

 In this paper we report the results of an all-sky search making use of  the  `weighted Hough'  method~\cite{Sintes:2006uc, :2007tda, Palomba:2005fp}.  The  `weighted Hough' was developed to improve the sensitivity of the `standard Hough' search~\cite{Abbott:2005pu, Krishnan:2004sv} and allows us to analyze data from multiple detectors, taking into account the different sensitivities.

The work presented here achieves improved sensitivity compared to previous Hough searches~\cite{Abbott:2005pu, :2007tda}  
 by splitting the run into two year--long portions and  requiring consistency between signal levels in the two separate years
for each candidate event, in addition to incorporating a $\chi^2$--test~\cite{SanchodelaJordana:2008dc}. 
This new pipeline is efficient at rejecting background, allowing us to lower the event threshold and achieve improved sensitivity. 
The parameter space searched in our analysis covers the frequency range 
$50<f<1000\,\mathrm{Hz}$ and the frequency time--derivative range
$-8.9\times10^{-10}<\dot{f}<0$ Hz/s. 
We detect no signals, so our results are presented as strain amplitudes $h_0$ excluded
at 95$\%$ confidence, marginalized over the above $\dot{f}$ interval.

Through the use of significant distributed computing resources \cite{Einstweb}, another search  \cite{Aasi:2012fw} has achieved better sensitivity on the same data as the search described here.
But the Einstein$@$Home production run on the second year of S5 LIGO data required about 9.5 months, used a total of 
approximately 25000 CPU (central processing unit) years \cite{Aasi:2012fw}, and required five weeks for the post-processing 
on a cluster with 6720 CPU cores.
The search presented in this paper used only 500 CPU months to process each of the two years of data, representing a computational cost more than two orders of magnitude smaller. 
This is also an order of magnitude smaller than the computational cost of the semi-coherent `PowerFlux' search reported in a
previous paper \cite{Abadie:2011wj}.
%

The significance of our analysis is through offering an
independent analysis to cross-check these results, and a method
 that allows the attainment of sensitivity
close to that of the Einstein$@$Home search at substantially reduced
computational burden. This technique will be particularly important in
the advanced LIGO and Virgo detector when applied to ``quick-look'' searches for nearby
sources that may have detectable electromagnetic counterparts.
Moreover, the Hough transform is more robust than other computationally efficient semi-coherent methods with respect to
noise spectral disturbances \cite{:2007tda} and phase modeling of the signal. In particular, it is also more robust than the 
Einstein$@$Home search to the non inclusion of second order frequency derivatives.

An important feature to note is that the sensitivity of the Hough search is proportional to $1/( N^{1/4} \sqrt{\Tcoh})$ or
$N^{1/4}/ \sqrt{\Tobs}$, assuming $\Tobs = N \Tcoh$, being $N$ the number of data segments coherently integrated over a time baseline $\Tcoh$ and combined using the  Hough transform over the whole observation time $\Tobs$, while for a coherent search over the whole observation time, the sensitivity is proportional to $1/ \sqrt{\Tobs}$.  This illustrates the lost of sensitivity introduced combining the different data segments incoherently but, of course, this is compensated by the lesser computational requirements of the semi-coherent method. 

For sufficiently short segments ($\Tcoh$ of the order of 30 minutes or less), the signal remains within a single Fourier frequency bin in each segment. In this case  a simple Fourier transform can be applied as a coherent integration method. As the segment duration 
$\Tcoh$ is increased, it becomes necessary to account  for signal modulations within each segment by computing the so-called
 ${\cal F}$-statistic \cite{Jaranowski:1998qm} over a grid in the space of phase evolution parameters, whose spacing decreases dramatically with time baseline $\Tcoh$. This results  in a significant increase in the computational requirements of the search and also limits the significant thresholds for data points selection and the ultimate sensitivity of the search.

The search presented here is based on 30 minute long coherent integration times, being this the reason for the significant reduction of the computational time compared to the Einstein$@$Home search \cite{Aasi:2012fw} in which the span of each segment was set equal to 25 hours. For an in-depth discussion on how to estimate and optimize the sensitivity of wide area searches for spinning neutron stars at a given computational cost, we refer the reader to \cite{Cutler:2005pn, Prix:2012yu, Wette:2011eu}. 

This paper is organized as follows: Section \ref{sec:data} briefly describes the LIGO interferometers and the data from 
LIGO's fifth science run. Section \ref{sec:waveform} defines the waveforms we seek and the associated assumptions we have made. 
In section  \ref{sec:houghtransform} we briefly review the Hough-transform method.
Section  \ref{sec:chi2veto} describes the  $\chi^2$ test  implemented for the analysis of the full S5 data.
Section  \ref{sec:pipeline} gives a detailed description of the search pipeline and results. 
Upper limit computations are provided in section \ref{sec:upperlimits}.
The study of some features related to the $\chi^2$-veto is presented in section \ref{sec:veto2}. 
Section \ref{sec:sensitivities} discusses variations,  further improvements and capabilities of alternative searches.
Section \ref{sec:end} concludes with a summary of the results.


\section{Data from the LIGO's fifth science run}
\label{sec:data}

During LIGO's fifth science run the LIGO detector network consisted of a 4-km interferometer  in Livingston, Louisiana (called L1) and two interferometers in Hanford, Washington, one a 4-km and another 2-km (H1 and H2, respectively). 
The fifth science run spanned a nearly two-year period of data acquisition. 
This run started at 16:00~UTC on November 4, 2005 at  Hanford  and at 16:00~UTC on November 14, 2005 at Livingston Observatory;  the run ended at 00:00~UTC on October 1, 2007.
During this run, all three LIGO detectors had displacement spectral amplitudes very near their design goals of $1.1\times 10^{ -19}$~m$\cdot$Hz$^{-1/2}$ in their most sensitive frequency band near $150$~Hz  for the 4-km detectors and, in terms of gravitational-wave strain, the H2 interferometer was roughly a factor of two less sensitive than the other two over most of the relevant band.
\begin{table}
\caption{\label{T.times}The reference GPS initial and final time  for the data collected during the  LIGO's fifth science run, together with the number of hours of data  used for the analysis. }
\begin{indented}
\item[]\begin{tabular}{@{}lllllll}
\br
& 1st year &&  & 2nd year &&  \\
Detector&start &end & hours &start & end& hours  \\
\mr
H1& 815410991& 846338742 &  5710  &    846375384 & 877610329 & 6295  \\
H2& 815201292 & 846340583 &  6097.5 &  846376386 & 877630716 &  6089\\
L1 & 816070323 & 846334700 &   4349&   846387978& 877760976 &  5316.5\\
\br
\end{tabular}
\end{indented}
\end{table}
%

The data were acquired and digitized at a rate of $16384$~Hz. Data acquisition was periodically interrupted by disturbances such as seismic transients (natural or anthropogenic), reducing the net running time of the interferometers. In addition, there were 1-2 week commissioning breaks to repair equipment and address newly identified noise sources. The resulting duty factors for the interferometers, defined as the fraction of the total run time when the interferometer was locked (i.e., all the interferometer control servos operating in their linear regime) and in its low configuration, were approximately $69\%$ for H1, $77\%$ for H2, and $57\%$ for L1 during the first eight months. A nearby construction project degraded the L1 duty factor significantly during this early period of the S5 run. By the end of the S5 run, the cumulative duty factors had improved to $78\%$ for H1, $79\%$ for H2, and $66\%$ for L1. 

In the paper  the data  from each of the three LIGO detectors is used to search for continuous gravitational wave signals.
 In table \ref{T.times} we provide the reference GPS initial and final times  for the data collected for each detector, together with the number of hours of data used for the analysis,
where each data segment used was required to contain at least 30 minutes of continuous interferometer operation.

\section{The waveform model}
\label{sec:waveform}

Spinning neutron stars may generate continuous gravitational waves (GW) through a variety of mechanisms. Independently of the specific mechanism, the emitted signal is a quasi-periodic wave whose frequency changes slowly during the observation time due to energy loss through gravitational wave emission, and possibly other mechanisms.
The form of the received signal at the detector is
\be
h(t)=F_+\left( t,\psi \right) h_+ \left( t \right) +  F_\times \left( t,\psi \right) h_\times \left( t \right)
\ee  
where $t$ is time in the detector frame, $\psi$ is the polarization angle of the wave and  $F_{+,\times}$ characterize the detector responses for the two orthogonal polarizations \cite{Bonazzola:1995rb, Jaranowski:1998qm}.   For an isolated quadrupolar gravitational-wave emitter, characterized by a rotating triaxial-ellipsoid mass distribution, 
 the individual components $h_{+,\times}$ 
have the form
\be
h_+ = h_0 \frac{1+\cos^2\iota}{2} \cos\Phi(t) \quad \textrm{and} \quad h_\times = h_0 \,\cos\iota\, \sin\Phi(t),
\ee
 where $\iota$ describes the inclination of the source's rotation axis to the line of sight,  $h_0$ is the wave amplitude and $\Phi(t)$ is the phase evolution of the signal. 
 For such a star, the gravitational wave frequency, $f$, is twice the rotation frequency and the amplitude  $h_0$ is given by
\begin{equation}
\label{eq:GWampl}
  h_0 = \frac{4\pi^2G}{c^4}\frac{I_{zz}f^2\varepsilon}{d},
\end{equation}
where $d$ is the distance to the star, $I_{zz}$ is the principal moment of inertia with respect to its spin axis, 
$\varepsilon$ the equatorial ellipticity of the star,  $G$ is Newton's constant and
$c$ is the speed of light.
 
 Note that the search method used in this paper is sensitive to periodic signals from  any type of isolated gravitational-wave source, though we present upper limits in terms of $h_0$.
 Because we use the Hough method, only the 
  instantaneous signal frequency  in the detector frame, $2\pi f(t) = d \Phi(t)/dt $,  needs to be calculated. This is given, to a very good approximation, by the non-relativistic Doppler expression: 
 \be
 \label{eq:Doppler}
 f(t)-\hat{f}(t) = \hat{f}(t) \frac{\textbf{v}(t)\cdot \textbf{n}}{c}, 
 \ee
 where  $\hat{f}(t)$ is the instantaneous signal frequency in the Solar System Barycenter (SSB), $\textbf{v}(t)$ is the detector velocity with respect to the SSB frame and $\textbf{n}$ is the unit-vector corresponding to the sky location of the source. In this analysis, we search for $\hat{f}(t)$ signals well described by a nominal frequency $f_0$ at the start time of the S5 run $t_0$ and a constant first time derivative $\dot f$, such that
 \be
 \hat{f}(t) = f_0+ \dot f (t-t_0).
 \ee
These equations ignore corrections to the time interval $t-t_0$ at the detector compared with that at the SSB and relativistic corrections. These corrections are negligible  for the search described here.

\section{The Hough transform}
\label{sec:houghtransform}


The Hough transform is a well known method for pattern recognition  that has been applied to the search for continuous gravitational waves.
In this case the Hough transform is used to find hypothetical signals  whose time-frequency evolution fits the pattern produced by
the Doppler modulation of the detected frequency, due to the Earth's rotational and orbital motion with respect to the Solar System Barycenter, and the time derivative of the frequency intrinsic to the source. 
 Further details can be found in  \cite{Krishnan:2004sv, Sintes:2006uc, Palomba:2005fp, Abbott:2005pu}; here we only give a brief summary.

The starting point for the Hough transform are $N$  short Fourier transforms (SFTs).
Each of these SFTs is digitized by setting a threshold $\rho_\th$ on the normalized power
\begin{equation} \label{eq:normpower}
\rho_k = \frac{2|\tilde{x}_k|^2}{\Tcoh S_n(f_k)} \,.
\end{equation}
Here  ${\tilde{x}_k}$ is the discrete Fourier transform of the data,
the frequency index $k$ corresponds to a physical frequency of $f_k= k/\Tcoh$,
$S_n(f_k)$ is the single sided power spectral density of the
detector noise and $\Tcoh$ is the time baseline of the SFT.
The $k^{th}$ frequency bin is selected if $\rho_k
\geq \rho_\th$, and rejected otherwise.  In this way, each SFT is replaced
by a collection of zeros and ones called a peak-gram.  This is the simplest method of selecting frequency bins, for which the  optimal choice of the threshold $\rho_\th$  is 1.6~\cite{Krishnan:2004sv}. Alternative conditions could be imposed \cite{doi:10.1142/S0218271800000438, Astone:2005fj, PhysRevD.66.102003},  that might be more robust against spectral disturbances.

For our choice, the probability that a frequency bin is selected is  $q = e^{-\rho_\th}$ for Gaussian noise and  $\eta$, given by
\begin{equation}
\label{eq:eta}
\eta = q\left\{1+\frac{\rho_\th}{2}\lambda_k +
\mathcal{O}(\lambda_k^2)  \right\}
\end{equation}
 is the corresponding probability in the presence of a signal. $\lambda_k$ is the signal to noise ratio (SNR) within a single SFT, and
for the case when there is no mismatch between the signal and the
template:
\begin{equation}
\label{eq:lambda}
\lambda_k = \frac{4|\tilde{h}(f_k)|^2}{\Tcoh S_n(f_k)}
\end{equation}
with $\tilde{h}(f)$ being the Fourier transform of the signal $h(t)$.

Several flavors of the Hough transform have been developed \cite{Krishnan:2004sv, Sintes:2006uc, Antonucci:2008jp} and used for different searches \cite{Abbott:2005pu, :2007tda, Collaboration:2009nc}.
The Hough transform is used to map points from the time-frequency plane of our data
(understood as a sequence of peak-grams) into the space of the source parameters.
 Each point in parameter space corresponds to a pattern in the time-frequency plane, 
and  the Hough number count $n$  is  the weighted sum of the ones and zeros,  $n_k^{(i)} $,
of the different peak-grams along this curve. For the `weighted Hough' this sum is computed as
\begin{equation}
\label{eq:2}
  n = \sum_{i=0}^{N-1} w^{(i)}_{k} n^{(i)}_{k}\,.
\end{equation}
where the  the choice of weights is optimal, in the sense of \cite{Sintes:2006uc}, if  defined as
\begin{equation} 
\label{eq:3}
  w^{(i)}_{k} \propto  \frac{1}{S^{(i)}_{k}}\left\{
    \left(F_{+1/2}^{(i)}\right)^2 +
    \left(F_{\times 1/2}^{(i)}\right)^2\right\},
\end{equation}
where $F_{+1/2}^{(i)}$ and $F_{\times 1/2}^{(i)}$ are the values of the beam pattern functions at the mid point of the $i^{th}$ SFT and 
%
%
are normalized  according to
\begin{equation}
 \label{eq:4}
  \sum_{i=0}^{N-1} w^{(i)}_{k} = N\,.
\end{equation}

The natural detection statistic is  the \textit{significance}  (or critical ratio) defined as:
\be
 s= \frac{n-\langle n\rangle}{\sigma} \, ,
 \ee
 where $\langle n\rangle$ and $\sigma$ are the expected mean and standard deviation
 for pure noise. 
 Furthermore,  
 the  relation between the significance and the  false alarm probability $\alpha$, in the Gaussian approximation
  \cite{Krishnan:2004sv},  is given by  
 \be
  \label{eq:sth}
 s_\th= \sqrt{2}\textrm{erfc}^{-1}(2\alpha)\,.
 \ee
 
%
%
%
%
%


\section{The $\chi^2$ veto}
\label{sec:chi2veto}

$\chi^2$ time-frequency discriminators are commonly used for gravitational  wave detection. Originally, they were designed  for broadband signals with a known waveform in a data stream \cite{Allen:2004gu}. But they can be adapted for narrowband continuous signals, as those expected from rapidly rotating neutron stars. The essence of these tests is to ``break up" the data (in time or frequency domain) and to see if the response in each chunk is consistent with what would be expected from the purported signal.

In this paper,  a chi-square test is implemented as a veto, in order to reduce the number of candidates in the analysis of the full S5 data.
The idea for this  $\chi^2$  discriminator is to 
 split the data into $p$  non-overlapping chunks,
each of them containing a certain number of SFTs $\{N_1, N_2,\ldots,N_p\}$, such that 
\be
\sum_{j=1}^p N_j =N \, ,
\ee
and analyze them separately, obtaining the  Hough
number-count $n_j$ which, for the same pattern across the different chunks,
 would then satisfy
\be
\sum_{j=1}^p n_j =n \, ,
\ee
where $n$ is the total number-count for a given point in parameter space.  
The  $\chi^2$ statistic will look along the different chunks to see if the number count  accumulates 
in a way that is consistent with the properties of the signal and 
the detector noise. Small values of $\chi^2$ are consistent with the hypothesis
that  the observed significance  arose from a detector output which was
a linear combination of Gaussian noise and the continuous wave signal. 
Large values of $\chi^2$ indicate either the signal did not match the template
or that the detector noise was non-Gaussian.
 
In the following  subsections we derive a $\chi^2$ discriminator for the different implementations of the Hough transform
and show how the veto curve was derived for LIGO S5 data.

\subsection{The standard Hough}
\label{sec:chi2SH}
In the simplest case in which all weights are set to unity,  the expected value and variance  of
the number count are
\be
 \langle n\rangle =N\eta \, ,
 \qquad  
 \sigma^2_n =N\eta(1-\eta) \, ,
 \ee
\be
 \langle n_j\rangle= N_j\eta= N_j\frac{\langle n\rangle}{N} \, ,
 \qquad
 \sigma^2_{n_j} =N_j\eta(1-\eta) \, .
\ee

Consider the $p$ quantities defined by
\be
\Delta n_j\equiv n_j-\frac{N_j}{N}n \, .
\ee
With this definition, it holds true that 
\be\label{eq:chi=0}
\langle \Delta n_j\rangle=0 \, ,
\qquad 
\sum_{j=1}^p \Delta n_j=0 \, , \qquad 
\langle  n_j n\rangle= \frac{N_j}{N} \langle n^2\rangle \,,
\ee
and the expectation value of the square of $\Delta n_j$ is
\be
 \langle (\Delta n_j)^2\rangle =\left( 1-\frac{N_j}{N}\right) N_j\eta(1-\eta) \, .
\ee
 Therefore we can define the $\chi^2$ discriminator statistic by
 \be
 \label{eq:chi3}
\chi^2(n_1, \ldots, n_p) =  
\sum_{j=1}^p {\frac{(\Delta n_j)^2}{\sigma^2_{n_j}} }=
\sum_{j=1}^p {\frac{\left(n_j-nN_j/N\right)^2}{N_j\eta(1-\eta)} }\, .
 \ee
 This corresponds to a 
$\chi^2$-distribution with $p-1$ degrees of freedom. To implement this discriminator, we need to 
measure, for each point in parameter space, the total number-count $n$, the
partial number-counts $n_j$ and  assume a constant value of $\eta=n/N$.

\subsection{The weighted Hough}
\label{sec:chi2WH}
In the case of the weighted Hough the result given by equation (\ref{eq:chi3})  can be generalized.
Let $I_j$ be the set of SFT indices for each
different $p$ chunks, thus  the mean and variance of the number-count become
\be
\langle  n_j \rangle= \sum_{i\in I_j} w_i\eta_i
\qquad
\langle  n \rangle=   \sum_{j=1}^p \langle  n_j \rangle
\qquad 
\sigma^2_{n_j} = \sum_{i\in I_j} w_i^2\eta_i(1-\eta_i) 
\ee
and we can define
\be
\Delta n_j\equiv n_j-n\frac{\sum_{i\in I_j} w_i\eta_i}{\sum_{i=1}^N w_i\eta_i} \, ,
\ee
 so that 
$\langle \Delta n_j\rangle=0$,
 $\sum_{j=1}^p \Delta n_j=0$. Hence, the $\chi^2$ discriminator would now be:
 \be
\chi^2 =  
\sum_{j=1}^p {\frac{(\Delta n_j)^2}{\sigma^2_{n_j}} } =
\sum_{j=1}^p {\frac{\left(n_j-n (\sum_{i\in I_j} w_i\eta_i)/(\sum_{i=1}^N w_i\eta_i)\right)^2}
{\sum_{i\in I_j} w_i^2\eta_i(1-\eta_i)} }\, . \label{eq:chi5}
 \ee
 
In a given search, we can compute the
 $\sum_{i\in I_j} w_i$, $\sum_{i\in I_j} w_i^2$ for each of the $p$ chunks, but the different $\eta_i$ values
 can not be measured from the data itself because they depend on the exact SNR for each single SFT
 as defined in equations (\ref{eq:eta}) and (\ref{eq:lambda}). For this reason, the discriminator we proposed is constructed by
 replacing   $\eta_i\rightarrow\eta^*$, where $\eta^*=n/N$. In this way, from 
equation (\ref{eq:chi5}) we get
 \be
 \label{eq:chi6}
\chi^2 \approx
\sum_{j=1}^p {\frac{\left(n_j-n (\sum_{i\in I_j} w_i)/N\right)^2}
{\eta^*(1-\eta^*) \sum_{i\in I_j} w_i^2 } }\, .
 \ee

In principle, one is free to choose the different $p$ chunks of data as one prefers, but it is reasonable to split the data into segments in such a way that they would contribute a similar relative contribution to the total number count. Therefore we split the SFT data in such a way that the sum of the weights in each block  satisfies
\be
\sum_{i\in I_j} \omega_{i} \approx \frac{N}{p} .
\ee

Further details and applications  of this  $\chi^2$ on LIGO S4 data can be found in ~\cite{SanchodelaJordana:2008dc}.

\subsection{The S5 $\chi^2$  veto curve}

We study the behavior of this  $\chi^2$ discriminator in order to characterize the  $\chi^2$-significance plane in the presence of signals and  derive empirically the veto curve.

For this purpose we use the full LIGO S5  SFT data, split  into both years, in the same way that is done in the analysis, and inject a  large number of Monte Carlo simulated continuous gravitational wave signals into the data, varying the amplitude, frequency, frequency derivative, sky location, as well as the nuisance parameters $\cos \iota$, $\psi$ and $\phi_0$ of the signals. Those injections are analyzed  with the multi-interferometer Hough code using the same grid resolution in parameter space as is used 
in the search.

\begin{figure}[h]
\begin{center}
\includegraphics[width=15pc]{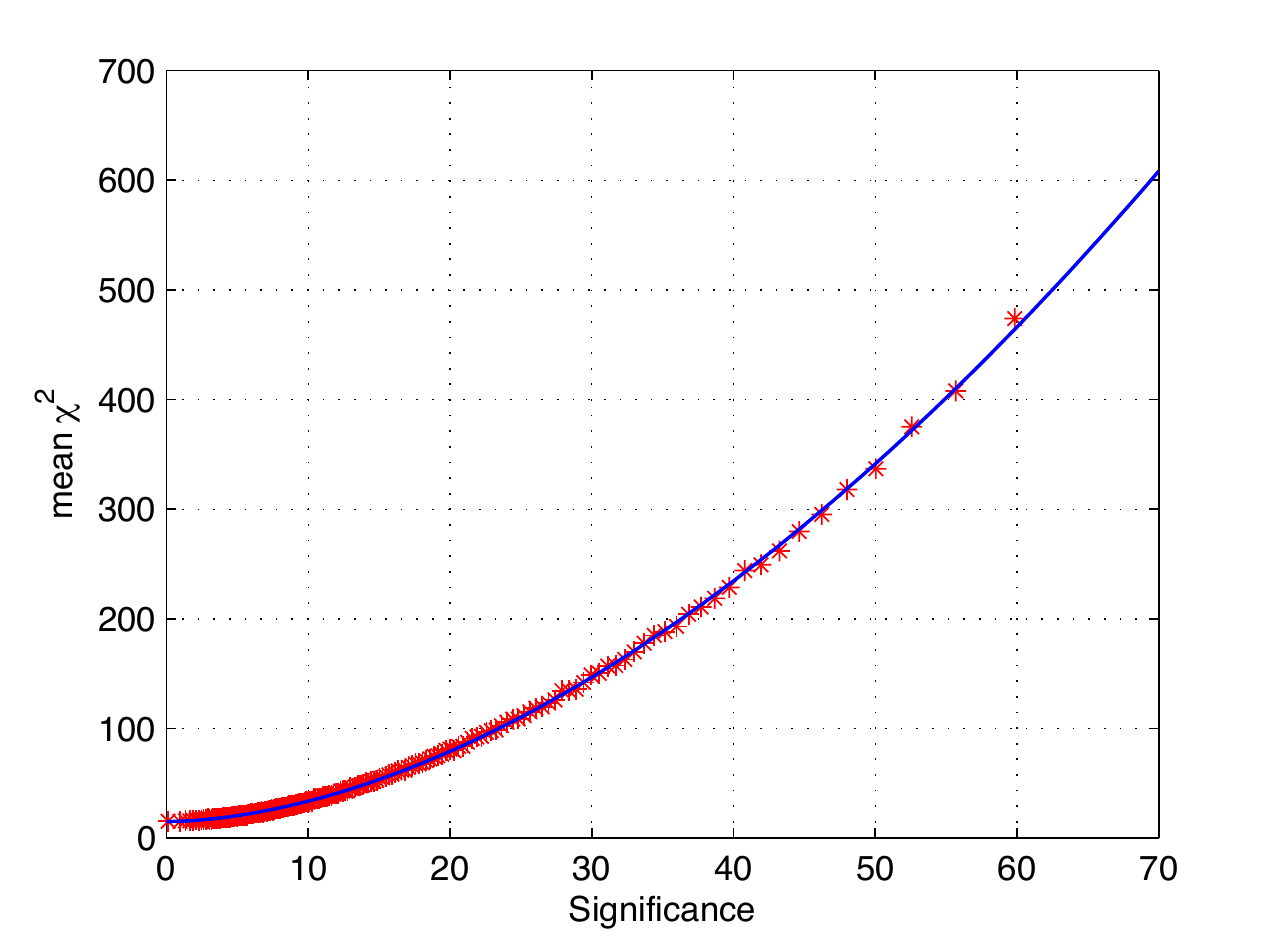}
\includegraphics[width=15pc]{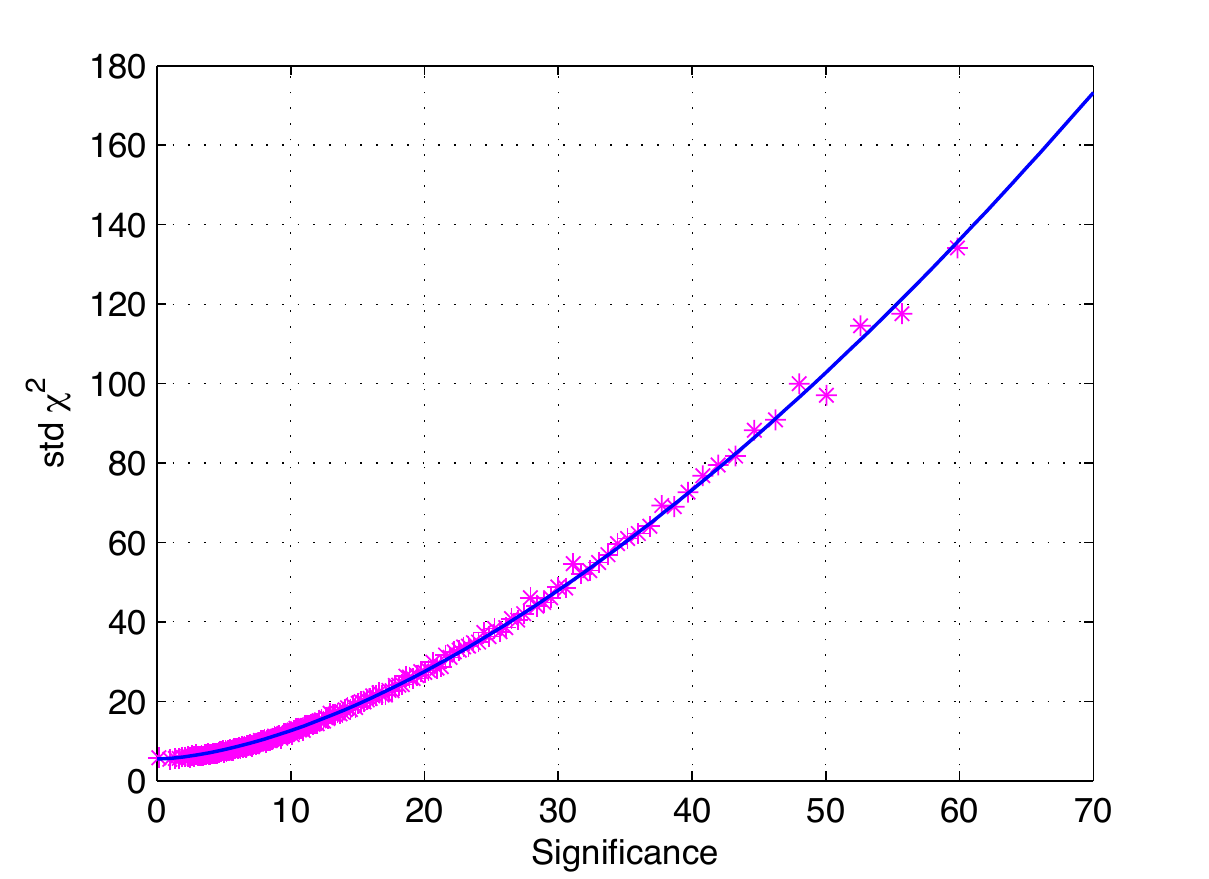}  \\
\includegraphics[width=32pc]{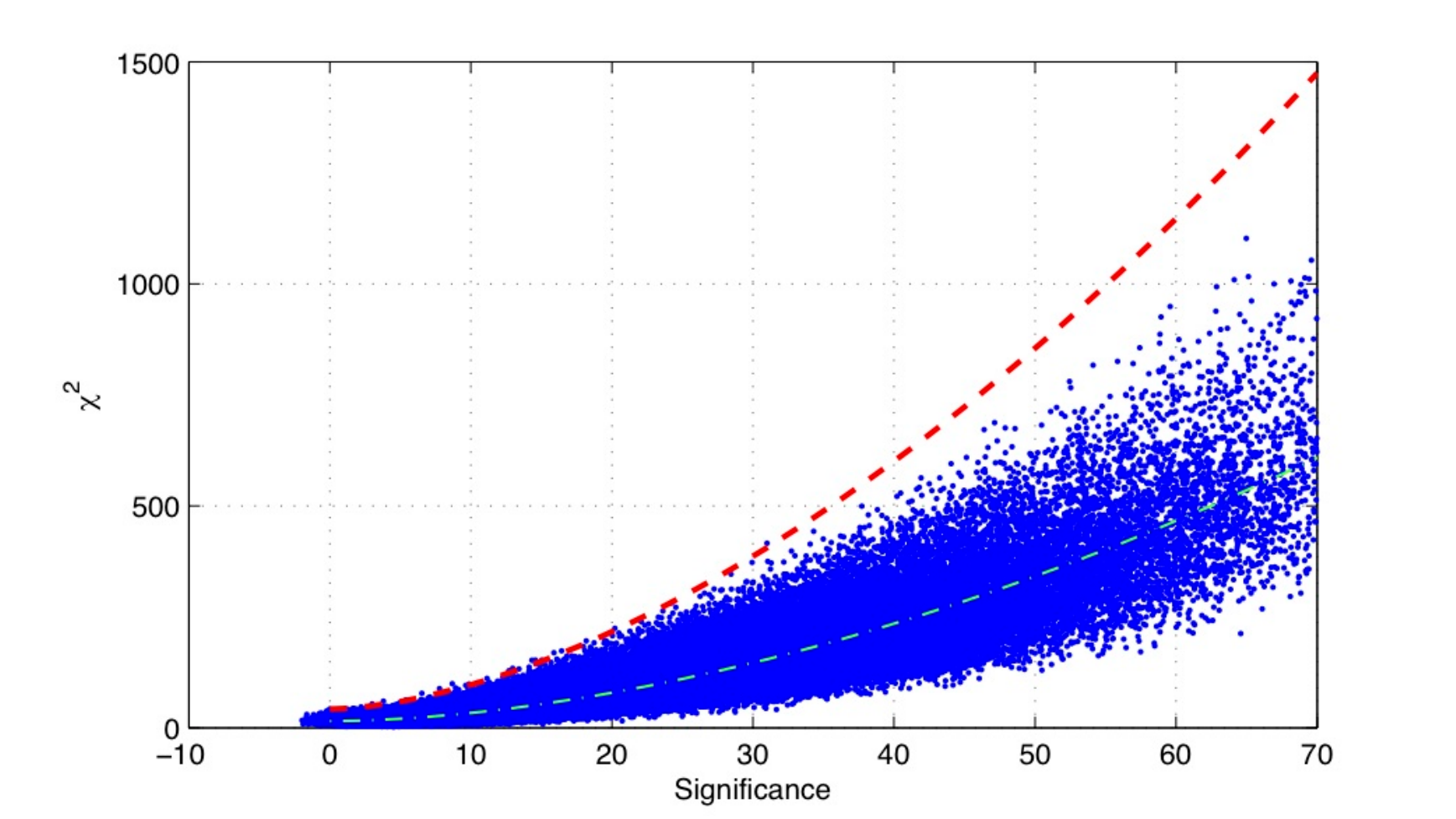} 
\end{center}
\caption{\label{fig:veto} (Top left) Mean value of the significance versus mean of the $\chi^2$ and the fitted power law curve for 177834  simulated injected signals. (Top right) Mean value of the significance versus mean $\chi^2$ standard deviation and the fitted power law curve. (Bottom)  Significance-$\chi^2$ plane for the injections, together with the fitted mean curve (dot-dashed line) and the veto curve (dashed line) corresponding to the mean $\chi^2$  plus five times its standard deviation.  }
\end{figure}

To characterize the veto curve,  nine 0.25Hz bands, spread in frequency and free of known large spectral disturbances have been selected. These are: 102.5Hz, 151Hz, 190Hz, 252.25Hz, 314.1Hz, 448.5Hz, 504.1Hz, 610.25Hz, and 710.25Hz. Monte Carlo injections in those bands  have been performed separately in the data from both years. Since the results were comparable for both years a single veto curved is derived.

In total 177834 injections are considered with a  significance  value lower than 70. The results of these injections in terms of ($s$, $\chi^2$) are presented in figure \ref{fig:veto}. The $\chi^2$ values obtained correspond to those by splitting the data in $p=16$ segments.

Then we proceed as follows: first we sort the points with respect to the significance, and we group them in sets containing 1000 points. For each set  we compute the  mean value of the significance, the mean of the $\chi^2$ and its standard deviation. With these reduced set of points we fit two  power laws $p-1+a \,s^c$ and $\sqrt{2p-2}+b \, s^d$ to the (mean $s$,  mean $\chi^2$) and  (mean $s$,  std $\chi^2$) respectively, obtaining the following coefficients (with $95\%$ confidence bounds):
\bea
a = 0.3123  & \quad & (0.305, 0.3195) \nonumber \\
c = 1.777     & \quad & (1.77, 1.783) \nonumber  \\
b = 0.1713  & \quad & (0.1637, 0.1789) \nonumber \\
d = 1.621    & \quad & (1.609, 1.633) \nonumber 
\eea

The veto curve we will use in this analysis corresponds to the mean curve plus five times the standard deviation
\be
 \bar\chi^2= p-1 + 0.3123 \, s^{1.777} + 5 (\sqrt{2p-2} + 0.1713 \, s^{1.621}).
\ee
This curve  vetoes  25 of  the 177834 injections considered with significance  lower than  70, that could translate into a  false dismissal rate of  0.014.
In figure  \ref{fig:veto} we  show the fitted curves and the $ \bar\chi^2$ veto curve compared to the result of the injections.


\section{Description of the all-sky search. }
\label{sec:pipeline}

\begin{figure}[t]
\begin{center}
\includegraphics[width=24pc]{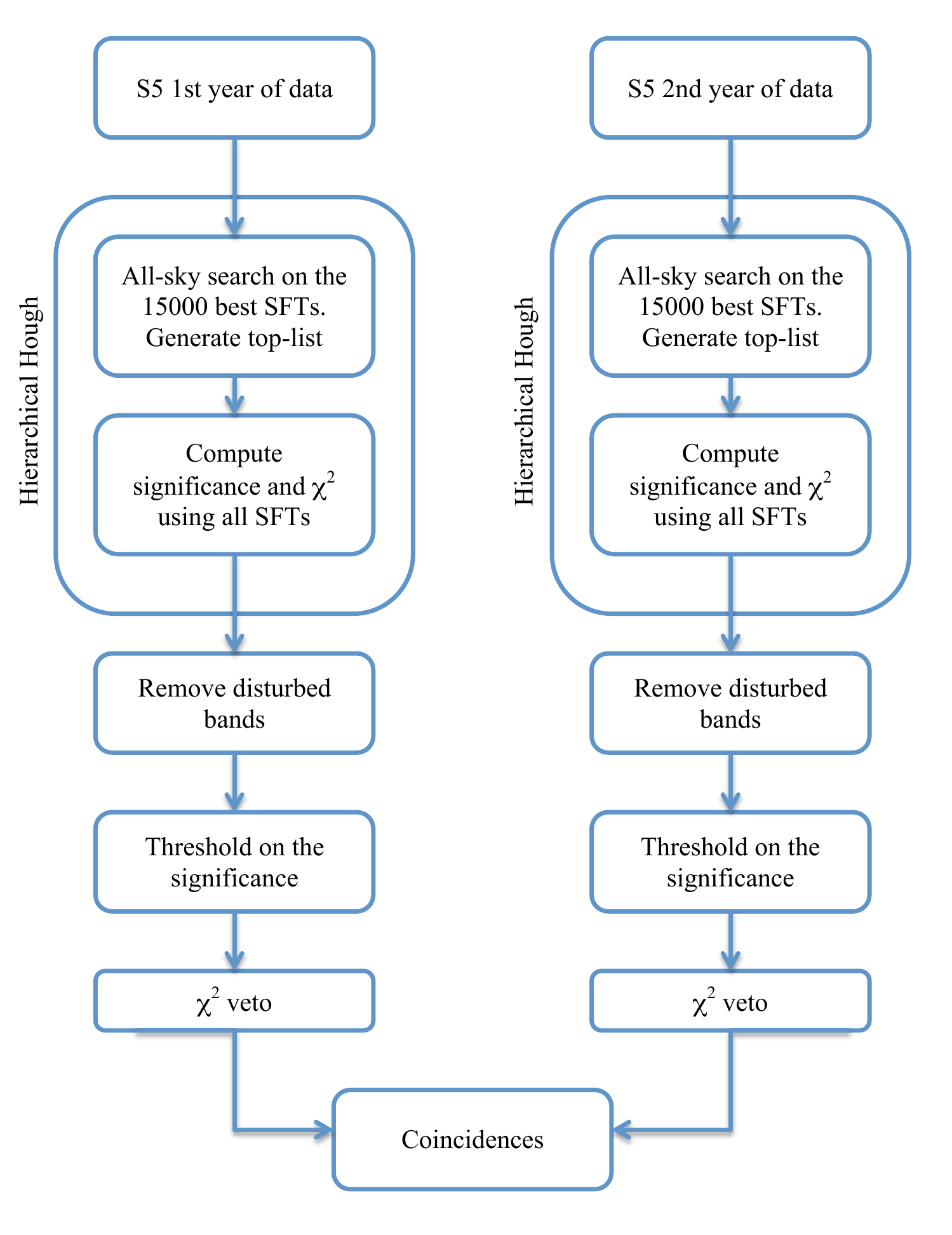}
\end{center}
\caption{\label{fig:pipe}Pipeline of the Hough search.  }
\end{figure}

In this paper, we use a new pipeline  to analyze the data from the fifth science run of the LIGO detectors to search for evidence of continuous gravitational waves, that might  be radiated by nearby  unknown  rapidly spinning
isolated neutron stars. Data from each of the three LIGO interferometers is used to perform the all-sky search. The key difference from previous searches is that, starting from $30~\mathrm{min}$ SFTs, 
we perform a multi-interferometer search analyzing separately the two years of the S5 run, and we study coincidences among the source candidates produced by the first and second years of data. Furthermore, we use a $\chi^2$ test adapted to the Hough transform searches to veto potential candidates. The pipeline is shown schematically  in figure  \ref{fig:pipe}.

A separate search was run for each successive 0.25 Hz band within
 the frequency range  $50\,$--$\,1000$~Hz
 and covering frequency  time derivatives in the range 
$-\sci{8.9}{-10}~\mathrm{Hz}~\mathrm{s}^{-1}$ to zero. 
We use a uniform grid spacing equal to the
size of a SFT frequency bin, $\delta f = 1/ \Tcoh= \sci{5.556}{-4}~\mathrm{Hz}$.
The resolution $\delta \dot{f}$  is given by the smallest value of $\dot{f}$ for which 
the intrinsic signal frequency does not drift by more than a  frequency bin during 
the observation time $\Tobs$ in the first year: 
$\delta \dot{f} = \delta f / \Tobs \sim \sci{1.8}{-11}~\mathrm{Hz}~\mathrm{s}^{-1}$.
This yields 51 spin-down values for each frequency. 
$\delta \dot{f}$  is fixed to the same value for the search on the first and the second year
 of S5 data.
The sky resolution, $\delta \theta$, is frequency dependent, 
as given by 
Eq.(4.14) of Ref.~\cite{Krishnan:2004sv}, that we increase by a factor 2. As explained in detail in Section V.B.1 of 
\cite{:2007tda}, the sky-grid spacing can be increased with a negligible loss in SNR, and  
for previous PowerFlux searches  \cite{:2007tda, :2008rg, Abadie:2011wj}
a factor 5 of increase was used in some frequency ranges to analyze LIGO S4 and S5 data.

The set of SFTs are generated  directly from the calibrated data stream, using 30-minute intervals of data for which the interferometer is operating in what is known as science mode. With this requirement, we search 32295 SFTs from the first year of S5 (11402 from H1, 12195 from H2 and 8698 from L1) and 35401 SFTs from the second year (12590 from H1, 12178 from H2 and 10633 from L1).

\subsection{A two-step hierarchical Hough search}
\label{subsec:hierarchicalHough}

The approach used to analyze each year of data is based on a two-step hierarchical search for continuous signals from isolated neutron stars. In both steps, the weighted Hough transform is used to find signals whose frequency evolution fits the pattern produced by the Doppler shift and the spin-down in the 
time-frequency plane of the data. The search is done by splitting the frequency range in $0.25$~Hz bands and using the SFTs from multiple interferometers. 

\begin{figure}[h]
\begin{center}
\includegraphics[width=32pc]{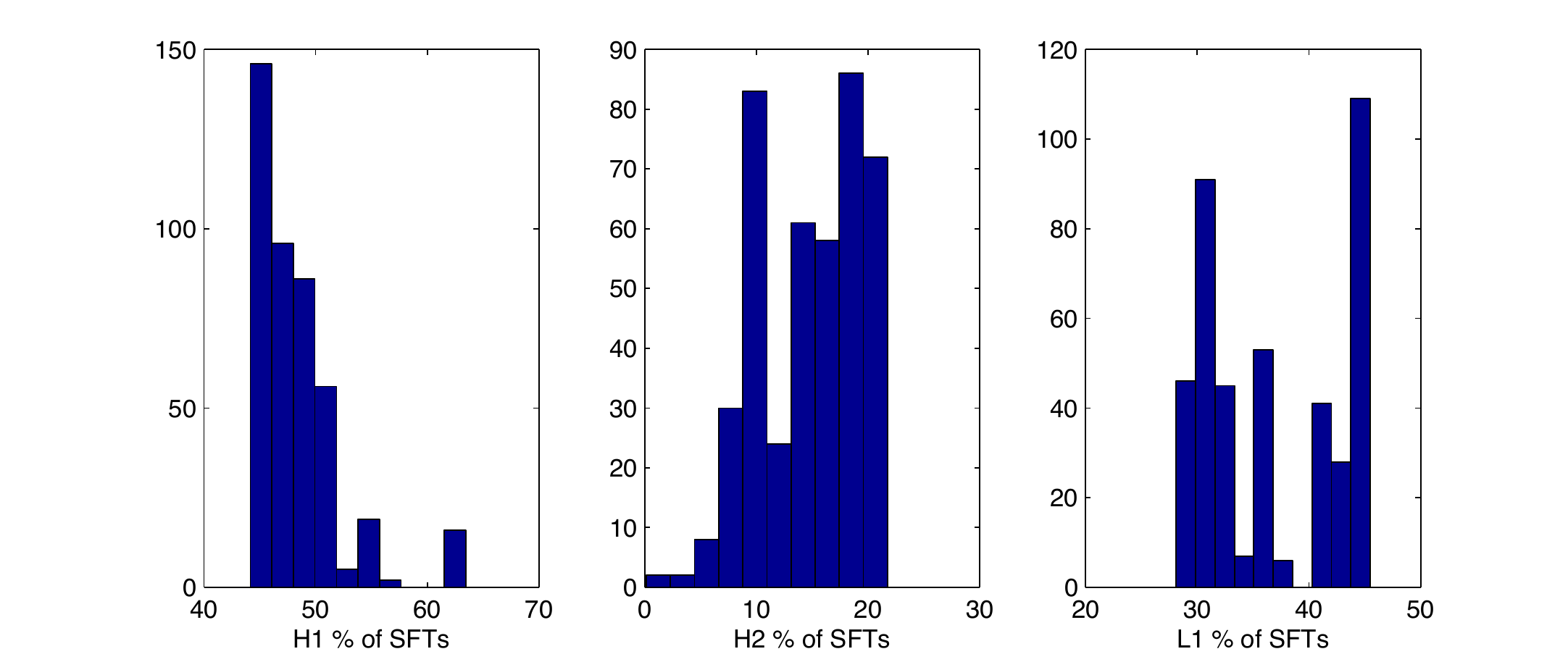}
\end{center}
\caption{\label{fig:SFTs} Histograms of the percentage of SFTs that each detector has contributed in the first stage to the all-sky search. These figures correspond to a 0.25 Hz band at 420 Hz for the first year of S5 data. The vertical axes are the number of sky-patches. }
\end{figure}

In the first stage, and for each  $0.25$~Hz  band, 
we break up the sky into smaller patches with frequency dependent size in order to use the  \textit{look up table} approach to compute the Hough transform, which greatly reduces the computational cost.
%
The \textit{look up table} approach benefits from the fact that, according to the Doppler expression (\ref{eq:Doppler}), the set of sky positions consistent with a given frequency bin  $f_k$ at a given time correspond to annuli on the celestial sphere centered on the velocity vector $\textbf{v}(t)$. In the \textit{look up table} approach, we precompute all the annuli for a given time  and a given search frequency mapped on the sky search grid. Moreover, it turns out that the mapped annuli are relatively insensitive to changes in frequency and can therefore be reused a large number of times. 
The Hough map is then constructed by selecting the appropriate annuli out of all the ones that have been found and adding them using the corresponding weights. A detailed description of the  \textit{look up table} approach with further details of implementation choices can be found in~\cite{Krishnan:2004sv}.

But limitations on the memory of the computers constrain the volume of data (i.e., the number of SFTs) that can be analyzed at once and the parameter space (e.g., size and resolution of the sky-patches and number 
of spin-down values) we can search over. For this reason, in this first stage, we select the best 15000 SFTs (according to the noise floor and the beam pattern functions) for each frequency band and  each sky-patch and apply the Hough transform on the selected data.
The size of the sky-patches ranges from $\sim 0.4~\mathrm{rad} \times 0.4~\mathrm{rad}$ at $50$~Hz to $\sim 0.07 ~\mathrm{rad} \times 0.07~\mathrm{rad}$ at $1$~kHz and we calculate the weights only for the center of each sky-patch. 
This was set in order to ensure that the memory usage will never exceed the 0.8GB and this search could run on the Merlin/Morgane dual compute cluster at the Albert Einstein Institute\footnote{http://gw.aei.mpg.de/resources/computational-resources/merlin-morgane-dual-compute-cluster}.
 A top-list keeping the best $1000$ candidates is produced for each $0.25$~Hz band for the all-sky search.

Figure \ref{fig:SFTs} shows the histograms of the percentage of SFTs that each detector contributes for the different sky locations for  a  band at  420~Hz for the first year of S5 data. At this particular frequency, the detector that contributes the most  is H1 between 44--64$\%$, giving the maximum contribution near the poles, L1 contributes between 28.1--45.5$\%$ with  its maximum around the equator, and H2 contributes at most $21.7\%$ of the SFTs. As shown in figure 1  in~\cite{delaJordana:2010za}, the maximum contribution of H2 corresponds to those sky regions  where L1 contributes the least.
If SFT selection had been based only upon the weights due to the noise floor, the H2 detector would not have contributed at all in this first stage.

In a second stage, we compute the $\chi^2$ value for all the  candidates in the top-list in each $0.25$~Hz band. 
This is done  by dividing the data into 16 chunks and summing weighted binary zeros or ones  along the expected path of the frequency evolution of a hypothetical periodic gravitational wave signal in the digitized time-frequency plane of our data. Since there are no computational limitations, we use the complete set of available SFTs from all three interferometers, and we also get a new value of the significance using all the data. In this way we reduce the mismatch of the template, since the number count is obtained  without the roundings introduced by the \textit{look up table} approach and  the weights are  computed for the precise sky location and not for the center of the corresponding patch. All these refinements contribute also  to a potential improvement of  sensitivity
when a threshold is subsequently applied to the recomputed significance (described below).

\begin{figure}[h]
\begin{center}
\includegraphics[width=32pc]{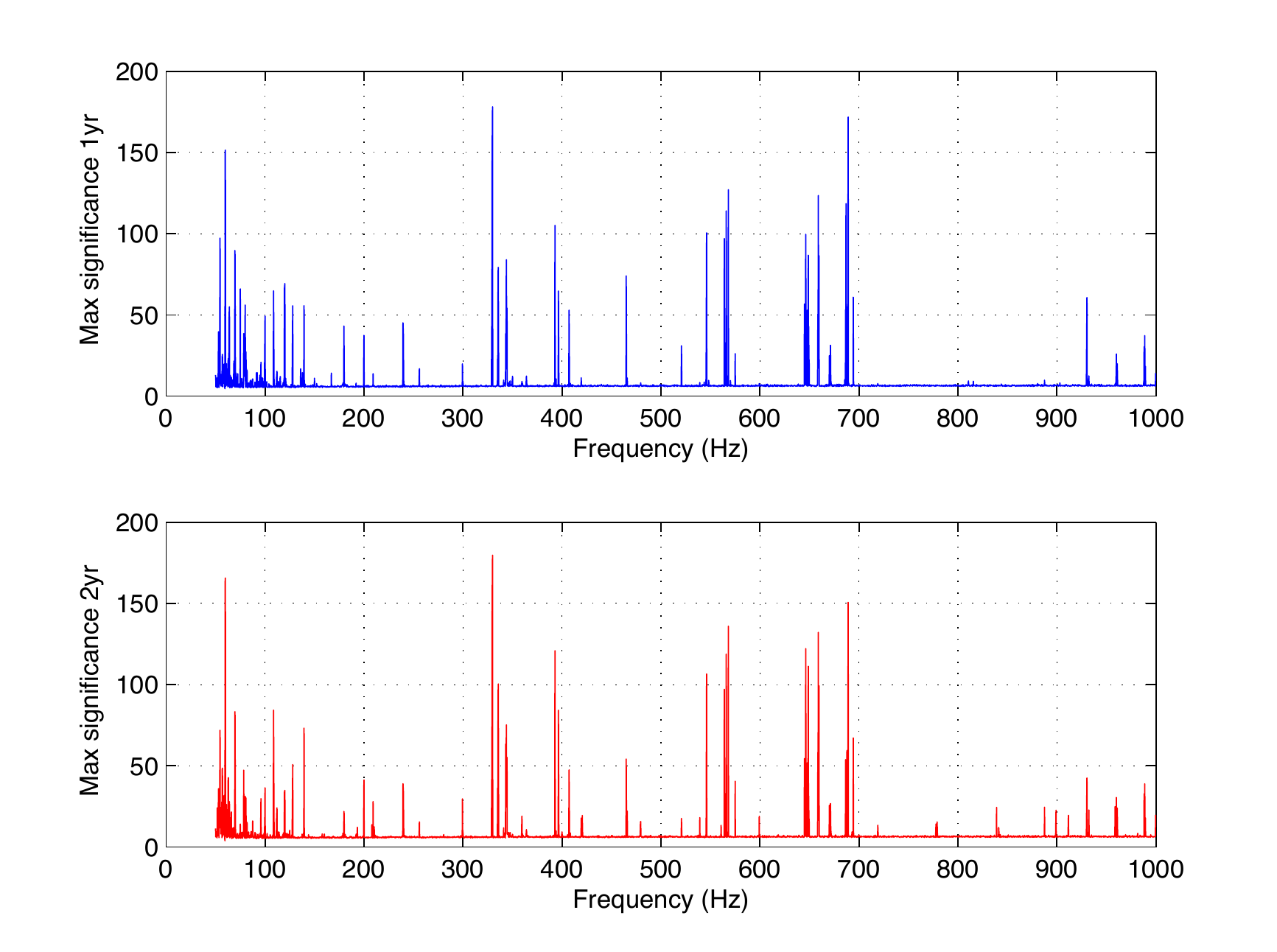}
\end{center}
\caption{\label{fig:MaxSig025} Maximum value of the significance for each $0.25$~Hz band for both years of LIGO S5 data. }
\end{figure}

Figure \ref{fig:MaxSig025} shows the maximum-significance value in each  $0.25$~Hz band obtained for the first and second years of S5 data.


\subsection{The post-processing}
\label{subsec:postpro}

After the  multi-interferometer Hough search is performed on each year of S5 data between 50 and 1000 Hz, a top list keeping the best 1000 candidates is produced for each 0.25 Hz band. This step yields $3.8\times10^6$ candidates for each year. 
The post-processing of these results has the following steps: 

\begin{enumerate} 
\item 
Remove those 0.25 Hz bands that are affected by power lines or violin modes.

A total of  96  bands are removed. These bands are given in table \ref{table:lines}.

\begin{table}
\caption{\label{table:lines} Initial frequency of the 0.25Hz bands excluded from the search.}
\begin{indented}
 \item[]\begin{tabular}{@{}c c}
   \br	
	Excluded Bands (Hz) & Description \\
	\mr
	$\left[ n60-0.25,n60+0.25 \right]$ $n$=1 to 16 & Power lines \\
	$\left[ 343.0 , 344.75  \right]$  & Violin modes \\
	$\left[ 346.5 , 347.75  \right]$  & Violin modes \\
	$\left[ 348.75, 349.25 \right]$  & Violin modes \\
	$\left[ 685.75 , 689.75 \right]$ & Violin mode harmonics\\
	$\left[ 693.0 , 695.5  \right]$   & Violin mode harmonics\\
	$\left[ 697.5 , 698.75 \right]$  & Violin mode harmonics\\
	\br
     \end{tabular}
  \end{indented}	
\end{table}

\begin{figure}[h]
\begin{center}
\includegraphics[width=32pc]{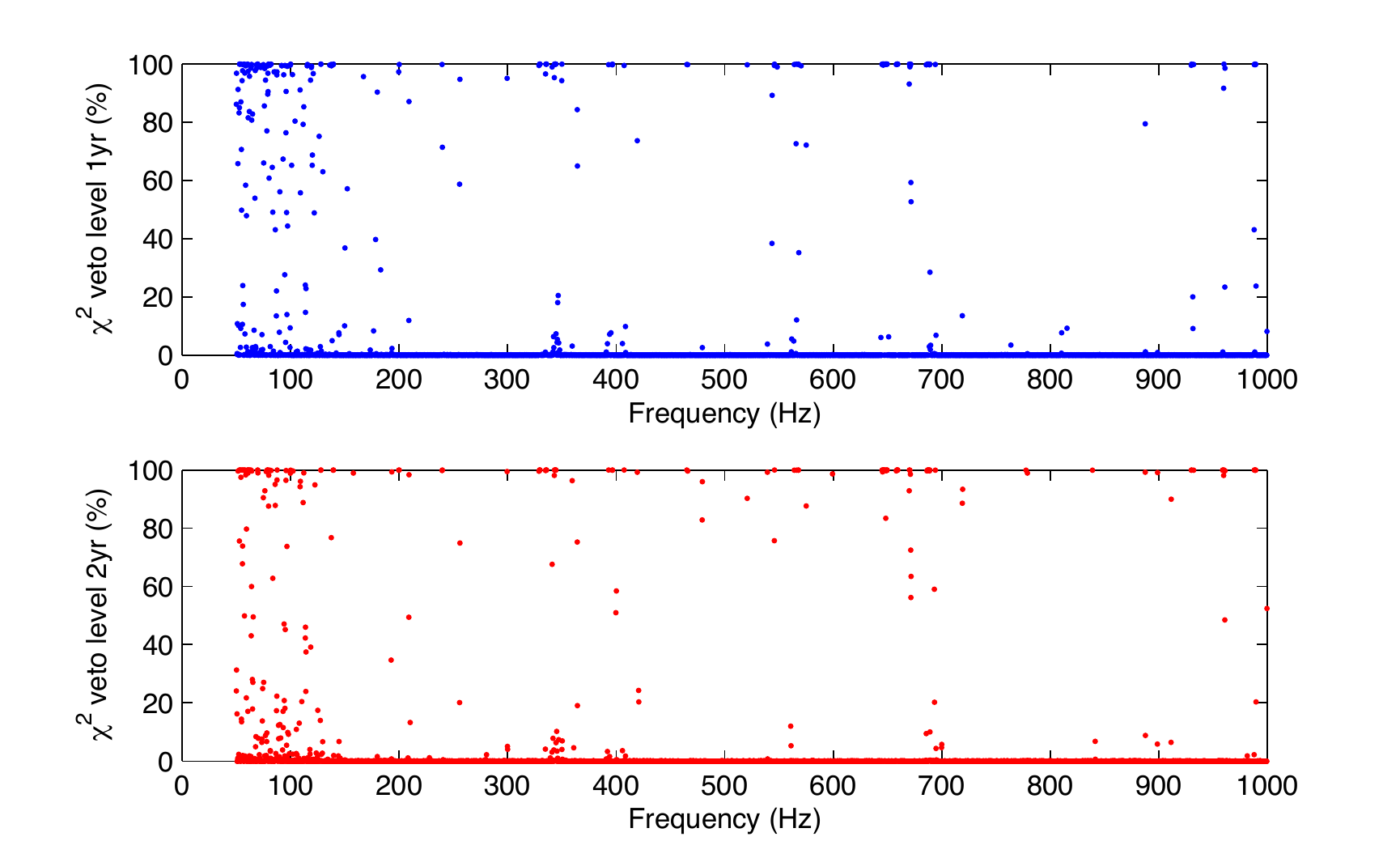}
\end{center}
\caption{\label{fig:Chi2Veto025} Percentage of the number of candidates vetoed due to a large $\chi^2$  value for each $0.25$~Hz band for both years of LIGO S5 data. }
\end{figure}

\item Remove all the  0.25~Hz bands for which the  $\chi^2$  vetoes more than a 95$\%$ of the elements in the top list.

Figure \ref{fig:Chi2Veto025} shows the   $\chi^2$ veto level for all the frequency bands for both years. With this criterion, 144 and 131 0.25~Hz bands would be vetoed for the first and second year of data respectively. These first two steps leave a total  3548 bands in which we search for coincidence candidates and  set upper limits; a total of 252  bands were discarded.

\begin{figure}[h]
\begin{center}
\includegraphics[width=15pc]{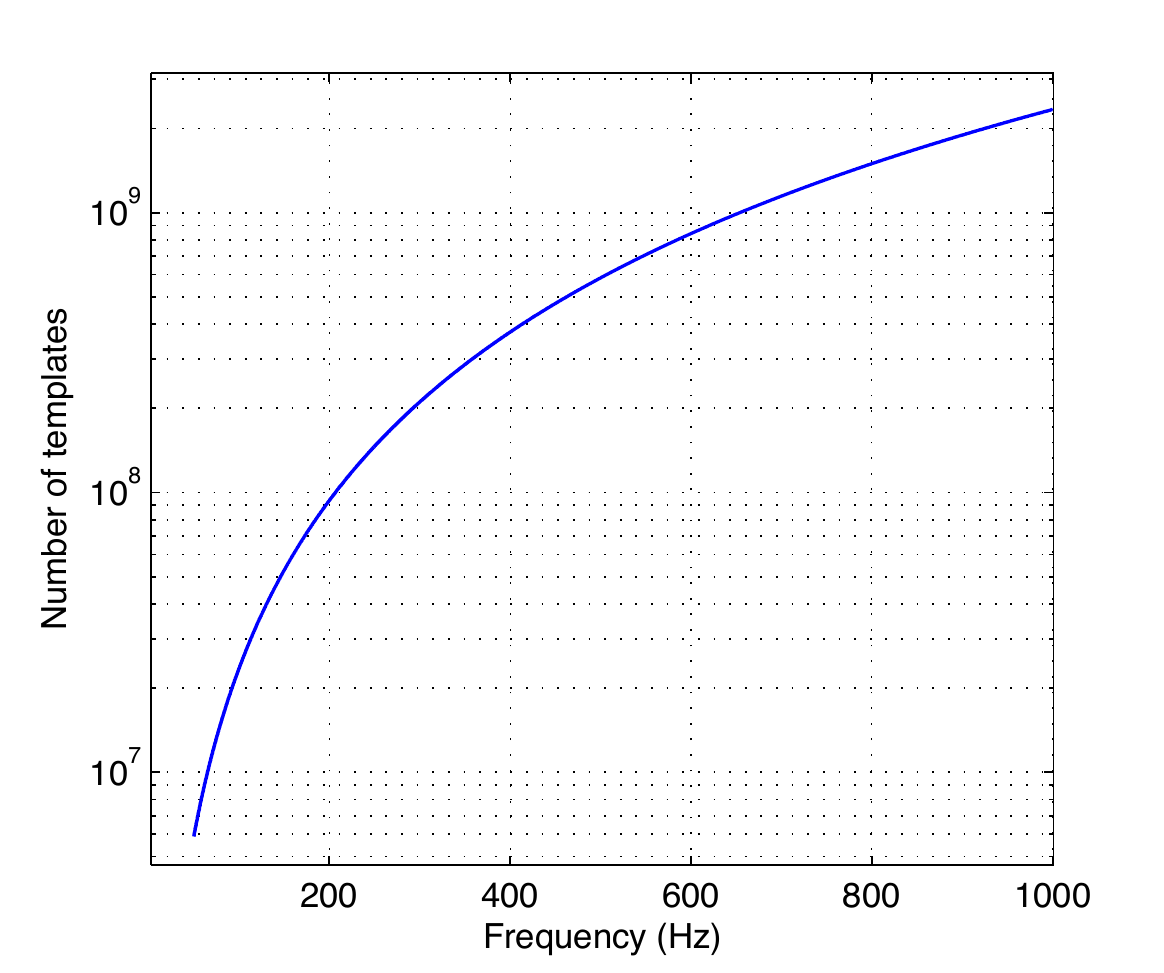}
\includegraphics[width=15pc]{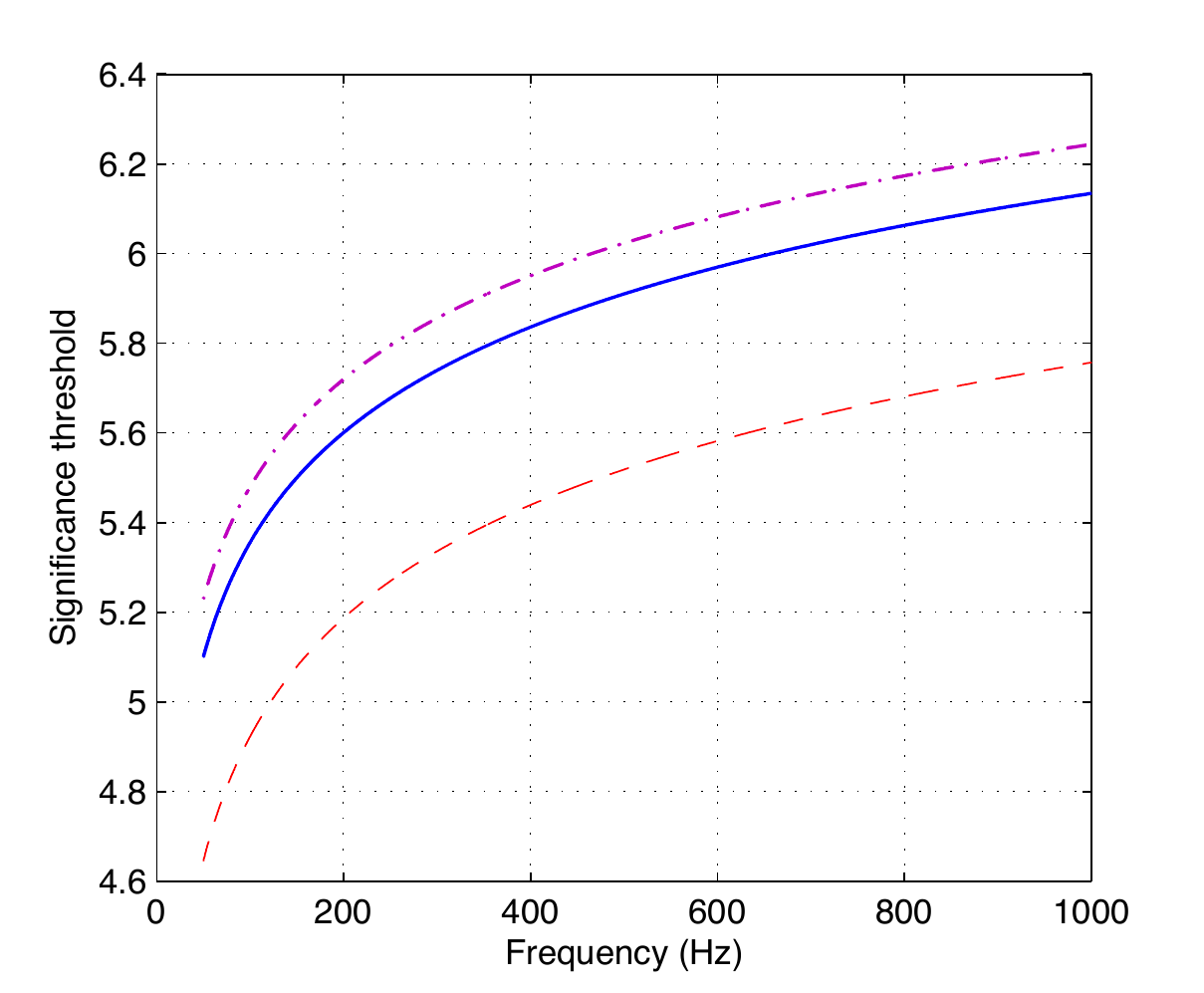}
\end{center}
\caption{\label{fig:templa} (Left) Number of templates analyzed in each  0.25~Hz band as a function of frequency. (Right) Significance threshold  for a false alarm level of  1/(number of templates) (solid line), compared to 10/(number of templates) (dashed line)  and  0.5/(number of templates) (dot-dashed line)  in each band.}
\end{figure}

\begin{figure}[h]
\begin{center}
\includegraphics[width=32pc]{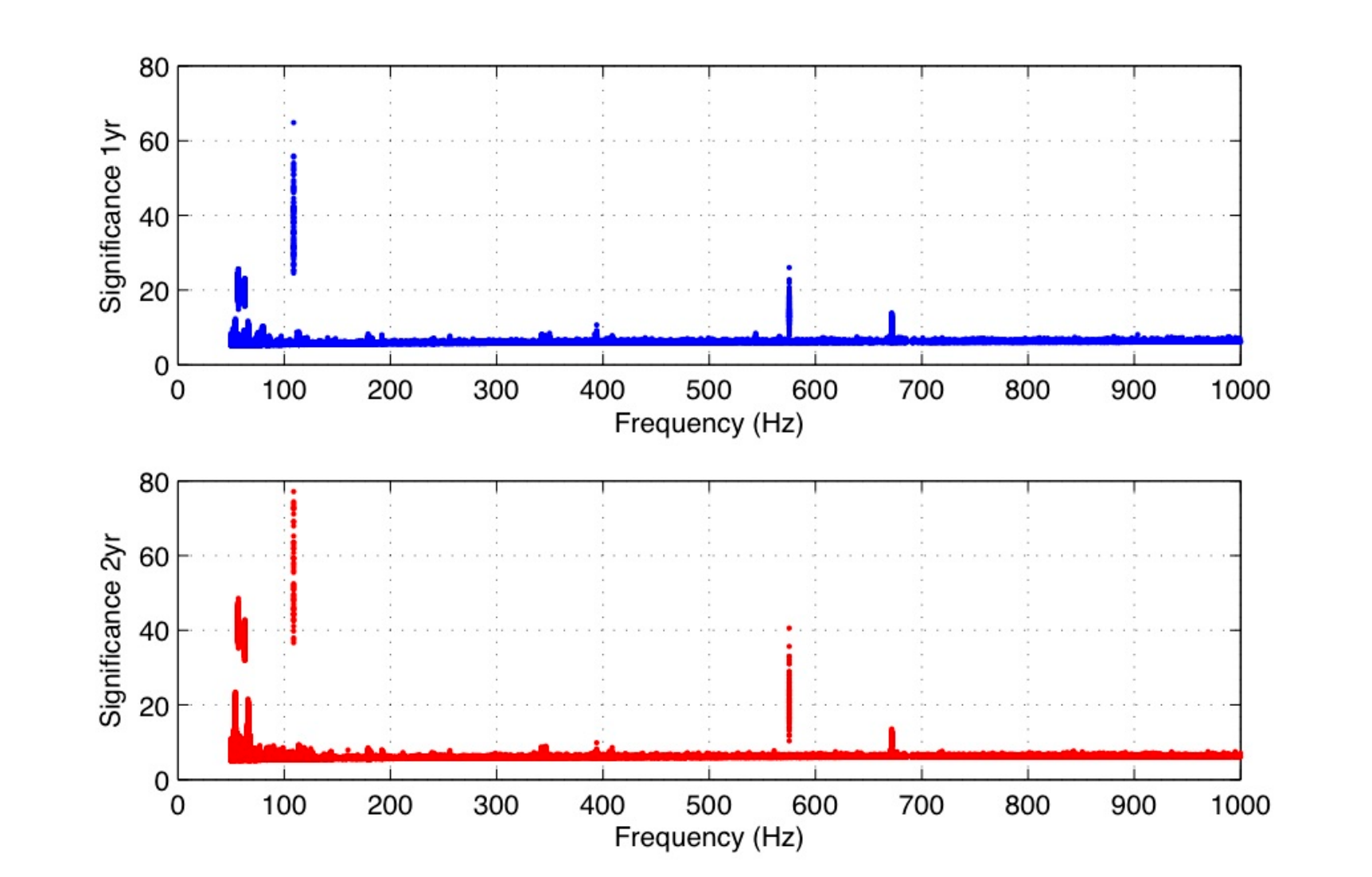}
\end{center}
\caption{\label{fig:candidates} Surviving candidates from both years after applying the $\chi^2$ veto and  setting a threshold in the significance. }
\end{figure}

\item Set a threshold on the significance.

Given the relation of the Hough significance and the Hough false alarm probability  (see equation (\ref{eq:sth})),  we  set a threshold on the candidate's significance that corresponds to a false alarm of 1/(number of templates) for each 0.25 Hz band. Figure  \ref{fig:templa} shows the value of this threshold at different frequencies.

\item Apply the  $\chi^2$ veto.

 From the initial  $3.8\times10^6$ elements in the top list for each year, after excluding the noisy bands, applying the $\chi^2$ veto and setting a threshold on the significance, the number of candidates remaining in the 3548 `clean' bands are  
 31427 for the 1st year and 50832 for the 2nd year. Those are shown in figure \ref{fig:candidates}.

\begin{figure}[h]
\begin{center}
\includegraphics[width=32pc]{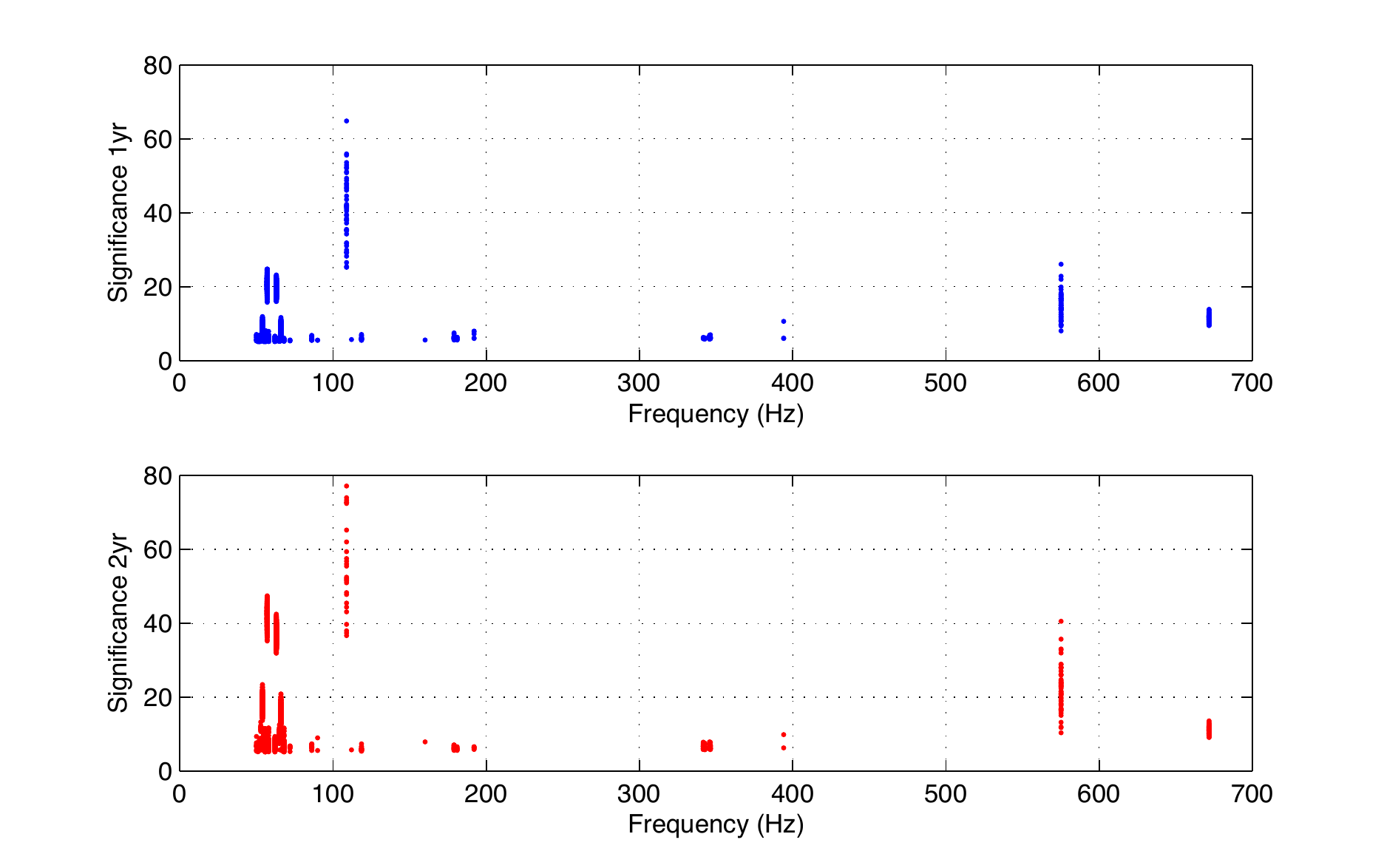}
\end{center}
\caption{\label{fig:coincidences} Significance of the coincidence candidates from the two years of LIGO S5 data. The upper and lower plots correspond to the first and second year respectively. }
\end{figure}

\item Selection of coincident candidates.

For each of the four parameters: frequency, spin-down and sky location, we set the coincidence window with a size equal to five times the grid spacing used in the search and centered on the values of the  candidates parameters. Therefore the coincidence window always  contains 625 cells in parameter space, with frequency-dependent size, according to the search grid. This window is computed for each of the candidates selected from the first year of data and  then we look for coincidences among the candidates of the second year, making sure to translate their  frequency  to  the reference time  of the starting time of the run, taking into account their spin-down  values.
%
Extensive analysis of software injected signals,  in different frequency bands, have been used to determine the size of this coincidence window. This was done by comparing the parameters of the most significant candidates of the search, using the same pipeline,  in both years of data.


With this procedure, we obtain 135728 coincidence pairs, corresponding to   5823 different  candidates of the first year that 
have coincidences with 7234 different ones of the second year. Those are displayed in figure \ref{fig:coincidences}.  All those candidates cluster in frequency in 34 groups. The most significant outlier  at 108.857 Hz corresponds to a simulated pulsar signal injected into the instrument as a test signal. The most significant events in each cluster are shown in tables \ref{table:candidates1y} and \ref{table:candidates2y}.
Notice how with this coincidence step the overall number of candidates has been reduced by a factor 5.4 for the first year and a factor 7.0 for the second year. Furthermore, without the coincidence step, the candidates are spread over all
frequencies, whereas the surviving coincident candidates are clustered in a few
small regions, illustrating the power of this procedure on real data.
    
 \begin{table}
\caption{\label{table:candidates1y} Summary of 1st year coincidence candidates, 
including the frequency band, the number of candidates in each cluster and
showing the details of the most significant  candidate in each of the 34 clusters.
Shown are the significance $s$, the  $\chi^2$ value, the detected frequency at the start of the run (SSB frame) $f_0$, 
the spin down $\dot f$, and the sky position (RA, dec).}
{\footnotesize
 \begin{tabular}{@{}r c r  r   r r r r r }
   \br	
	& Band (Hz) & Num. & $s$  & $\chi^2$ & $f_0$ (Hz) & $\dot f$ (Hz s$^{-1}$)  &RA (rad) &dec (rad)  \\
	\mr
    1 & 50.001 - 50.003 & 12 &  7.103 & 64.115  & 50.0022 &  0 & -1.84  & 0.69 \\ 
2 & 50.997 - 51.004 & 7 &  5.431 & 57.322  & 51.0028 &  0 & -2.73  & 0.94 \\ 
3 & 52.000 - 52.016 & 44 &  6.886 & 54.108  & 52.0139 &  -21.4e-11 & -1.99  & -0.08 \\ 
4 & 52.786 - 52.793 & 6 &  6.094 & 40.089  & 52.7911 &  0 & 0  & 1.33 \\ 
5 & 53.996 - 54.011 & 1136 & 11.880 & 60.583  & 54.0039 &  -10.7e-11 & 2.68  & -1.21 \\ 
6 & 54.996 - 55.011 & 82 &  6.995 & 48.289  & 55.0067 &  0 & 3.00  & -0.47 \\ 
7 & 55.749 - 55.749 & 7 &  5.940 & 37.902  & 55.7489 &  0 & -0.23  & 1.12 \\ 
8 & 56.000 - 56.016 & 167 &  8.085 & 66.130  & 56.0056 &  -7.1e-11 & 0.20  & 0.22 \\ 
9 & 56.997 - 57.011 & 1370 & 24.751 & 175.459  & 57.0028 &  -12.5e-11 & -2.10  & 1.40 \\ 
10 & 58.000 - 58.015 & 89 &  7.901 & 61.360  & 58.0128 &  -35.7e-11 & -2.32  & 0.43 \\ 
11 & 61.994 - 62.000 & 195 &  6.597 & 26.445  & 61.9989 &  0 & 0.98  & 0.21 \\ 
12 & 62.996 - 63.009 & 1324 & 23.188 & 131.619  & 63.0028 &  -8.9e-11 & 0  & -1.40 \\ 
13 & 64.996 - 65.000 & 101 &  6.252 & 52.374  & 64.9994 &  -5.3e-11 & -0.17  & 0.03 \\ 
14 & 65.378 - 65.381 & 3 &  5.586 & 29.000  & 65.3806 &  -16.1e-11 & -2.17  & 1.15 \\ 
15 & 65.994 - 66.013 & 919 & 11.662 & 35.587  & 65.9994 &  -7.1e-11 & 0.15  & 1.38 \\ 
16 & 67.006 - 67.006 & 1 &  5.801 & 11.950  & 67.0056 &  -17.8e-11 & -1.13  & -0.20 \\ 
17 & 67.993 - 68.009 & 39 &  6.061 & 44.319  & 68.0017 &  -14.3e-11 & -1.55  & 0.66 \\ 
18 & 72.000 - 72.000 & 4 &  5.604 & 17.668  & 72.0000 &  -1.8e-11 & 1.57  & -1.11 \\ 
19 & 86.002 - 86.024 & 14 &  6.786 & 47.054  & 86.0150 &  -17.8e-11 & 1.78 & 1.09 \\ 
20 & 90.000 - 90.000 & 2 &  5.554 & 58.338  & 90.0000 &  -3.6e-11 & 1.54  & -1.05 \\ 
21 & 108.857 - 108.860 & 50 & 64.850 & 552.161  & 108.8570 &  0 & 3.10  & -0.60 \\ 
22 & 111.998 - 111.998 & 1 &  5.673 & 46.493  & 111.9980 &  -7.1e-11 & -0.54  & 1.18 \\ 
23 & 118.589 - 118.613 & 18 &  7.072 & 52.181  & 118.5990 &  -57.1e-11 & 2.90  & -0.46 \\ 
24 & 160.000 - 160.000 & 1 &  5.571 & 18.847  & 160.0000 &  0 & 1.56  & -1.15 \\ 
25 & 178.983 - 179.026 & 21 &  7.483 & 43.380  & 179.0010 &  -3.6e-11 & -1.59  & 1.17 \\ 
26 & 181.000 - 181.038 & 8 &  6.309 & 13.174  & 181.0170 &  -8.9e-11 & -1.12 & -0.89 \\ 
27 & 192.000 - 192.002 & 5 &  7.976 & 41.042  & 192.0000 &  -1.8e-11 & -1.51  & 1.17 \\ 
28 & 341.763 - 341.765 & 3 &  6.292 & 39.340  & 341.7630 &  -1.8e-11 & -1.63 & 1.21 \\ 
29 & 342.680 - 342.684 & 5 &  6.113 & 36.893  & 342.6800 &  -3.6e-11 & 1.47  & -1.14 \\ 
30 & 345.721 - 345.724 & 17 &  6.835 & 38.540  & 345.7230 &  -12.5e-11 & -1.07  & 1.44 \\ 
31 & 346.306 - 346.316 & 9 &  6.973 & 30.298  & 346.3070 &  -3.6e-11 & 1.32  & -1.13 \\ 
32 & 394.099 - 394.100 & 3 & 10.617 & 95.063  & 394.1000 &  0 & -1.58  & 1.16 \\ 
33 & 575.163 - 575.167 & 57 & 26.058 & 146.929  & 575.1640 &  -1.8e-11 & -2.53  & 0.06 \\ 
34 & 671.728 - 671.733 & 101 & 13.878 & 132.197  & 671.7290 &  0 & 1.54  & -1.17 \\ 
\br
     \end{tabular}
}
\end{table}

           \begin{table}
\caption{\label{table:candidates2y} Summary of 2nd year coincidence candidates, showing the details of the most significant candidate in each of the 34 clusters.}
 {\footnotesize
\begin{tabular}{@{}r c r  r   r r r r r }
  \br	
	& Band (Hz) &  Num. & $s$   & $\chi^2$ & $f_0$ (Hz) & $\dot f$ (Hz s$^{-1}$)  & RA (rad) & dec (rad)  \\
	\mr
1 & 50.001 - 50.004 & 10 &  9.345 & 77.005  & 50.0006 &  -5.3e-11 & 1.25  & -1.05 \\ 
2 & 50.993 - 51.003 & 16 &  7.106 & 17.670  & 51.0011 &  -1.8e-11 & -1.95  & 1.32 \\ 
3 & 52.000 - 52.012 & 38 &  8.800 & 30.888  & 52.0094 &  -17.8e-11 & -2.17  & -0.08 \\ 
4 & 52.784 - 52.792 & 14 & 13.269 & 128.801  & 52.7911 &  -7.1e-11 & 0.77  & 1.31 \\ 
5 & 53.996 - 54.006 & 1380 & 23.431 & 69.473  & 54.0033 &  -14.3e-11 & -1.92 & 1.31 \\ 
6 & 54.995 - 55.008 & 405 & 11.554 & 73.521  & 55.0056 &  -5.3e-11 & -2.98  & 0.13 \\ 
7 & 55.748 - 55.749 & 46 &  7.102 & 34.151  & 55.7489 &  -1.8e-11 & -0.10  & 0.13 \\ 
8 & 56.000 - 56.008 & 161 & 11.478 & 64.077  & 56.0006 &  -1.8e-11 & 1.58  & -1.13 \\ 
9 & 56.996 - 57.004 & 1478 & 47.422 & 412.817  & 56.9989 &  0 & -0.31  & 1.43 \\ 
10 & 58.000 - 58.008 & 166 & 11.675 & 65.764  & 58.0000 &  0 & 1.59  & -1.13 \\ 
11 & 61.993 - 62.001 & 400 &  9.361 & 56.108  & 61.9989 &  0 & -1.16  & 1.02 \\ 
12 & 62.997 - 63.004 & 1138 & 42.432 & 360.615  & 63.0039 &  -5.3e-11 & -1.51  & 1.41 \\ 
13 & 64.994 - 65.000 & 288 & 12.557 & 119.361  & 64.9978 &  0 & 0.61  & -1.14 \\ 
14 & 65.377 - 65.378 & 2 &  5.777 & 34.580  & 65.3772 &  -14.3e-11 & -1.84  & 1.03 \\ 
15 & 65.995 - 66.006 & 1162 & 20.864 & 67.221  & 66.0022 &  -5.3e-11 & -1.51  & 1.40 \\ 
16 & 66.999 - 66.999 & 5 &  6.207 & 19.546  & 66.9989 &  -17.8e-11 & -1.08  & 0.03 \\ 
17 & 67.994 - 68.007 & 199 & 11.682 & 82.662  & 68.0006 &  -1.8e-11 & 1.65  & -1.14 \\ 
18 & 71.999 - 72.000 & 10 &  6.802 & 45.727  & 72.0000 &  -1.8e-11 & 1.41  & -1.14 \\ 
19 & 86.002 - 86.013 & 15 &  7.368 & 54.707  & 86.0094 &  -10.7e-11 & 2.10  & 0.75 \\ 
20 & 89.999 - 90.000 & 4 &  9.008 & 86.486  & 90.0000 &  0 & 1.54  & -1.19 \\ 
21 & 108.857 - 108.858 & 27 & 77.157 & 1613.230  & 108.8580 &  -5.3e-11 & 2.99  & -0.71 \\ 
22 & 111.996 - 111.996 & 1 &  5.775 & 59.202  & 111.9960 &  -7.1e-11 & -0.37  & 1.00 \\ 
23 & 118.579 - 118.589 & 19 &  7.398 & 69.372  & 118.5820 &  -41.0e-11 & 2.79  & -0.68 \\ 
24 & 160.000 - 160.000 & 2 &  7.898 & 15.079  & 160.0000 &  0 & 1.60  & -1.17 \\ 
25 & 178.984 - 179.014 & 24 &  7.089 & 33.229  & 179.0010 &  -10.7e-11 & -1.79  & 1.35 \\ 
26 & 180.998 - 181.019 & 8 &  6.587 & 28.605  & 181.0180 &  -28.5e-11 & -0.39  & 1.07 \\ 
27 & 191.999 - 192.001 & 8 &  6.582 & 45.301  & 192.0000 &  -3.6e-11 & 1.51  & -1.19 \\ 
28 & 341.762 - 341.764 & 19 &  7.859 & 71.736  & 341.7630 &  -3.6e-11 & -1.63  & 1.17 \\ 
29 & 342.677 - 342.680 & 19 &  7.569 & 45.255  & 342.6790 &  -14.3e-11 & 1.47  & -1.11 \\ 
30 & 345.718 - 345.721 & 12 &  7.890 & 44.570  & 345.7200 &  -12.5e-11 & -0.91  & 1.38 \\ 
31 & 346.303 - 346.309 & 14 &  7.803 & 45.756  & 346.3070 &  0 & 1.46  & -1.21 \\ 
32 & 394.099 - 394.100 & 2 &  9.877 & 95.232  & 394.1000 &  0 & -1.58  & 1.16 \\ 
33 & 575.163 - 575.165 & 53 & 40.576 & 415.830  & 575.1640 &  -1.8e-11 & -2.53  & 0.06 \\ 
34 & 671.729 - 671.732 & 87 & 13.600 & 116.261  & 671.7320 &  -1.8e-11 & 1.59  & -1.15 \\ 
\br
     \end{tabular}
     }
\end{table}

    \end{enumerate}

Noise lines were identified by previously performed searches (\cite{:2008rg, Collaboration:2009nc, Aasi:2012fw, Abadie:2011wj, Abbott:2009ws}) as well as the search described in this paper. Several techniques were used to identify the causes of outliers, including  the calculation of the coherence between the interferometer output channel and physical environment monitoring channels and the computation of high resolution spectra.  A dedicated analysis code ``FScan" \cite{Coughlin}  was also created specifically for identification of instrumental artifacts. Problematic noise lines were recorded and monitored throughout S5. 

In addition, a number of particular checks  were performed on the coincidence outliers, including: a
 detailed study of the full top-list results,   for those 0.25Hz bands where the candidates were found --in order to 
check  if  candidates are more dominant in a given year, or if they  cluster in certain regions of parameter space;  and  a second search using the data of the two most  sensitive detectors, H1 and L1 separately --in order to see if artifacts could be associated to  a given detector, consistent with the observed spectra.

All of the 34 outliers were investigated and were all traced  to instrumental  artifacts or hardware injections (see details in table \ref{table:description}). Hence the search did not reveal any true continuous gravitational wave signals.
   
       \begin{table}
\caption{\label{table:description} Description of the coincidence outliers, together with the maximum value of the significance in both years.}
  {\footnotesize
\begin{tabular}{@{}r c r r  c}
   \br	
	& Bands (Hz) & s 1y & s 2y & Comment \\
	\mr
1 & 50.001 - 50.004 &  7.103 &  9.345  & L1 1 Hz Harmonic from control/data acquisition system \\ 
2 & 50.993 - 51.004 &  5.431 &  7.106  & L1 1 Hz Harmonic from control/data acquisition system  \\ 
3 & 52.000 - 52.016 &  6.886 &  8.800  & L1 1 Hz Harmonic from control/data acquisition system  \\ 
4 & 52.784 - 52.793 &  6.094 & 13.269  &  Instrumental line in H1 \\ 
5 & 53.996 - 54.011 & 11.880 & 23.431  & Pulsed heating sideband on 60 Hz mains \\ 
6 & 54.995 - 55.011 &  6.995 & 11.554  & 1 Hz Harmonic from control/data acquisition system \\ 
7 & 55.748 - 55.749 &  5.940 &  7.102  & Instrumental line in L1\\ 
8 & 56.000 - 56.016 &  8.085 & 11.478  &  L1 1 Hz Harmonic from control/data acquisition system \\ 
9 & 56.996 - 57.011 & 24.751 & 47.422  & Pulsed heating sideband on 60 Hz mains\\ 
10 & 58.000 - 58.015 &  7.901 & 11.675  & L1 1 Hz Harmonic from control/data acquisition system \\ 
11 & 61.993 - 62.001 &  6.597 &  9.361  & L1 1 Hz Harmonic from control/data acquisition system \\ 
12 & 62.996 - 63.009 & 23.188 & 42.432  & Pulsed heating sideband on 60 Hz mains \\ 
13 & 64.994 - 65.000 &  6.252 & 12.557  & L1 1 Hz Harmonic from control/data acquisition system \\ 
14 & 65.377 - 65.381 &  5.586 &  5.777  & Instrumental line in L1 -- member of offset 1Hz comb  \\ 
15 & 65.994 - 66.013 & 11.662 & 20.864  & Pulsed sideband on 60 Hz mains  \\ 
16 & 66.999 - 67.006 &  5.801 &  6.207  &  L1 1 Hz Harmonic from control/data acquisition system  \\ 
17 & 67.993 - 68.009 &  6.061 & 11.682  & 1 Hz Harmonic from control/data acquisition system  \\ 
18 & 71.999 - 72.000 &  5.604 &  6.802  &  1 Hz Harmonic from control/data acquisition system \\ 
19 & 86.002 - 86.024 &  6.786 &  7.368  &  Instrumental line in H1 \\ 
20 & 89.999 - 90.000 &  5.554 &  9.008  & Instrumental line in H1 \\ 
21 & 108.857 - 108.860 & 64.850 & 77.157  & Hardware injection of simulated signal (ip3) \\ 
22 & 111.996 - 111.998 &  5.673 &  5.775  & 16 Hz harmonic from data acquisition system \\ 
23 & 118.579 - 118.613 &  7.072 &  7.398  & Sideband of mains at 120 Hz \\ 
24 & 160.000 - 160.000 &  5.571 &  7.898  & 16 Hz harmonic from data acquisition system \\ 
25 & 178.983 - 179.026 &  7.483 &  7.089  & Sideband of mains at 180 Hz \\ 
26 & 180.998 - 181.038 &  6.309 &  6.587  & Sideband of mains at 180 Hz  \\ 
27 & 191.999 - 192.002 &  7.976 &  6.582  & 16 Hz harmonic from data acquisition system\\ 
28 & 341.762 - 341.765 &  6.292 &  7.859  & Sideband of suspension wire resonance in H1 \\ 
29 & 342.677 - 342.684 &  6.113 &  7.569  & Sideband of suspension wire resonance in H1 \\ 
30 & 345.718 - 345.724 &  6.835 &  7.890  & Sideband of suspension wire resonance in H1 \\ 
31 & 346.303 - 346.316 &  6.973 &  7.803  & Sideband of suspension wire resonance in H1 \\ 
32 & 394.099 - 394.100 & 10.617 &  9.877  & Sideband of calibration line at 393.1 Hz in H1 \\ 
33 & 575.163 - 575.167 & 26.058 & 40.576  & Hardware injection of simulated signal (ip2) \\ 
34 & 671.728 - 671.733 & 13.878 & 13.600  & Instrumental line in H1 \\ 
	\br
     \end{tabular}
}
\end{table}

%

\section{Upper limits estimation and astrophysical reach}
\label{sec:upperlimits}

The analysis of the Hough search presented here has not identified any convincing continuous gravitational wave signal. Hence, we proceed to set upper limits on the maximum intrinsic  gravitational wave strain $h_0$  that is consistent with our observations for a population of signals described by an isolated triaxial rotating neutron star. 

As in the previous S2 and S4 searches \cite{Abbott:2005pu, :2007tda}, we set a population-based frequentist upper limit, assuming random positions in the sky, in the gravitational wave frequency range $[50,1000]$ Hz and with spin-down values in the range
$-\sci{8.9}{-10}~\mathrm{Hz}~\mathrm{s}^{-1}$ to zero. The rest of the nuisance parameters, $\cos \iota$, $\psi$ and $\phi_0$, are assumed to be uniformly distributed. As commonly done in all-sky, all-frequency  searches, the upper limits are given in different frequency sub-bands,  here chosen to be  0.25 Hz wide. Each upper limit is based on the most significant event from each year in its 0.25 Hz band. 
Our goal is to find the value of $h_0$ (denoted   $h_0^{95\%}$) such that $95\%$ of the signal injections at this amplitude would be recovered by our search and are more significant than the most significant candidate from the actual search in that band,  thus providing the $95\%$ confidence all-sky upper limit on $h_0$.

\begin{figure}[h]
\begin{center}
\includegraphics[width=32pc]{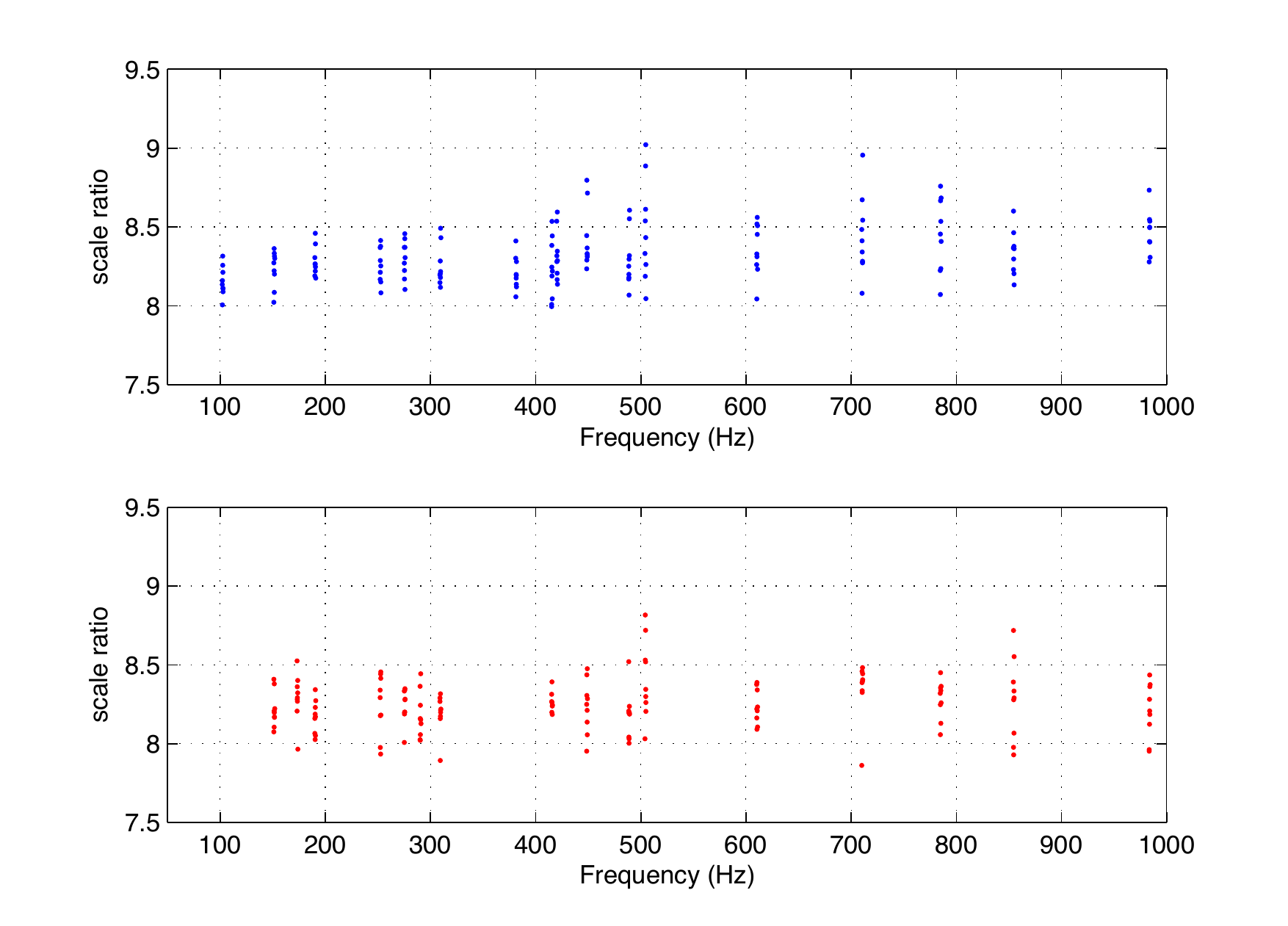}
\end{center}
\caption{\label{fig:scalefactor} Ratio of the upper limits measured by means of Monte-Carlo injections in the multi-interferometer Hough search to the quantity  $h_0^{95\%}/C$ as defined in Equation~(\ref{eq:ul}). 
The top figure corresponds to the first year of S5 data and the bottom one to the second year.
The comparison is performed by doing 500 Monte-Carlo injections for 10 different amplitude in several small frequency bands. 153 and 144 frequency bands have been used for the first and second year respectively.  }
\end{figure}

\begin{figure}[h]
\begin{center}
\includegraphics[width=34pc]{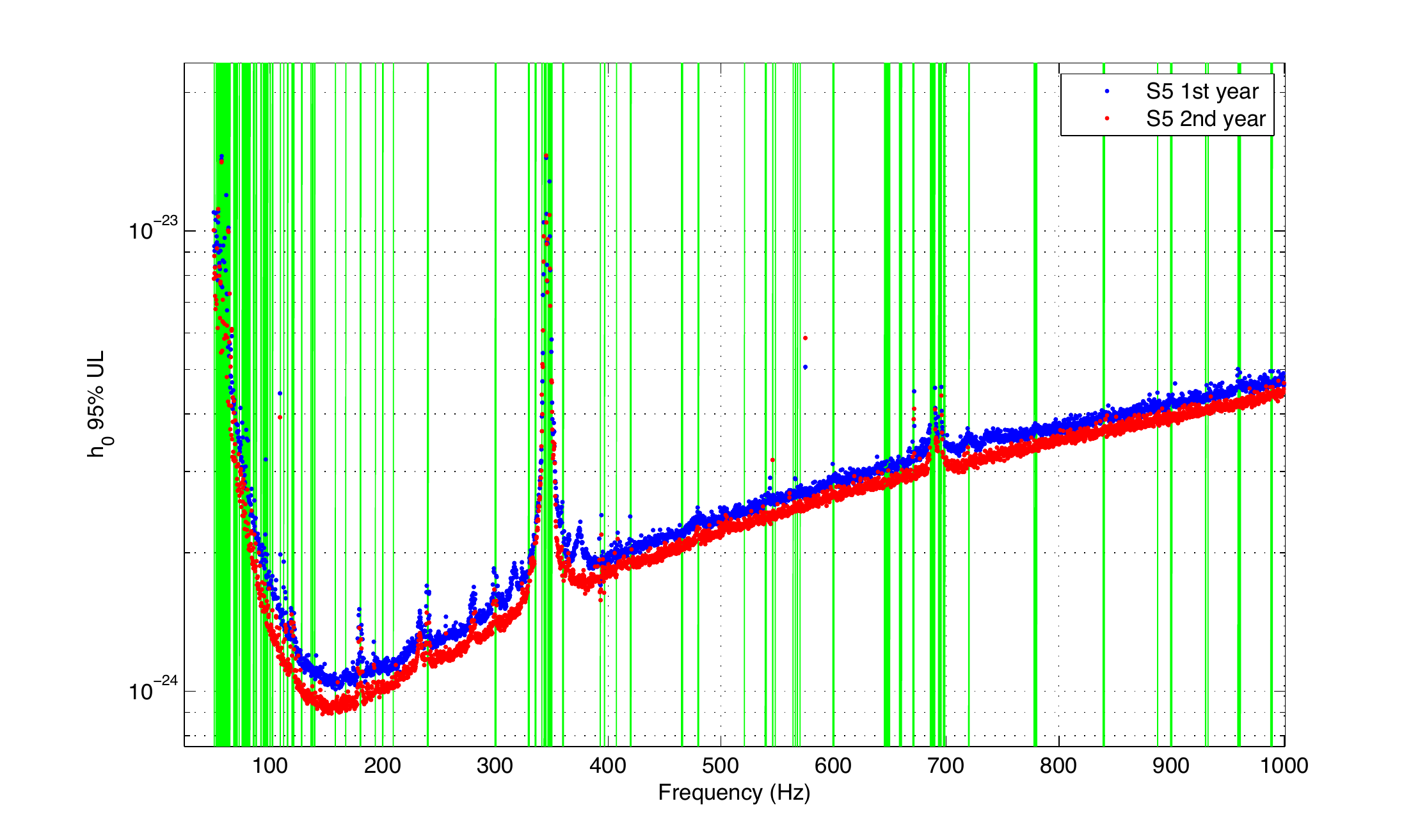}
\end{center}
\caption{\label{fig:UL} The $95\%$ confidence all-sky upper limits on $h_0$ from the hierarchical Hough multi-interferometer search together with excluded frequency bands. The best upper limits correspond to $1.0\times 10^{-24}$ for the S5 first year in the $158-158.25$~Hz band, and $8.9 \times 10^{-25}$ for the S5 second year in the $146.5-146.75$~Hz band. }
\end{figure}

Our procedure for setting upper-limits uses partial Monte Carlo signal injection studies, using the same search pipeline as described above, together with an analytical sensitivity estimation. 
As in the previous S4 Hough search  \cite{:2007tda},
upper limits  can be computed accurately without extensive Monte Carlo simulations. Up to a constant factor $C$, that depends on the grid resolution in parameter space,  they are given by

\begin{equation}
  \label{eq:ul}
  h_0^{95\%} = C \left( \frac{1}{
      \sum_{i=0}^{N-1}(S^\srchTemplateInd_{\iSubSupInd})^{-2}}\right)^{1/4}\sqrt{\frac{
      \S}{\Tcoh}}  \,. 
\end{equation}
where
\be
\label{eq:S}
 \S= \textrm{erfc}^{-1}(2\alpha_\H) + \textrm{erfc}^{-1}(2\beta_\H)  \,,
 \ee
 $S_i$ is the average value of the single sided power spectral noise density  of the $i^{th}$ SFT in the corresponding frequency sub-band,
$\alpha_\H$  is the  false alarm  and $\beta_\H$ the false dismissal probability.

The utility of this fit is that having determined the value of $C$ in a small frequency range, it can be extrapolated to cover the full bandwidth without performing any further Monte Carlo simulations. 
 Figure~\ref{fig:scalefactor} shows the value of the constant $C$  for a number of $0.1\,$Hz frequency bands. More precisely, this is the ratio of the upper limits measured by means of Monte-Carlo injections in the multi-interferometer Hough search to the quantity  $h_0^{95\%}/C$ as defined in equation~(\ref{eq:ul}). The value of $\S$  is computed using the false alarm $\alpha_\H$ corresponding to the observed loudest event, in a given frequency band, and  a false dismissal rate  $\beta_\H=0.05$, in correspondence to the desired confidence level of 95$\%$, i.e., $\S\rightarrow s^*/ \sqrt{2} + \textrm{erfc}^{-1}(0.1)$, where $s^*$ is the highest significance value in the frequency band. 
 This yields a scale factor $C$ of $8.32\pm0.19$ for the first year and $8.25\pm0.16$ for the second year of S5.
 With these values we proceed to set the upper limits for all the frequency bands. 
%
The validity of equation~(\ref{eq:ul}) was studied in  \cite{:2007tda} using LIGO  S4 data. In that paper upper limits were measured for each 0.25 Hz frequency band from 100 Hz to 1000 Hz using  Monte Carlo injections and compared with those prescribed by this analytical approximation. Such comparison study showed that the values obtained using 
equation~(\ref{eq:ul}) have an error smaller than $5\%$ for bands free of large instrumental disturbances. 
%
For an in-depth study of how to analytically estimate the sensitivity of wide parameter searches for gravitational-wave pulsars, we refer the reader to \cite{Wette:2011eu}.

 The 95$\%$ confidence all-sky upper limits on $h_0$ from this multi-interferometer  search for each year of S5 data are shown
in figure \ref{fig:UL}.  The best upper limits correspond to $1.0\times 10^{-24}$ for the first year of S5 in the $158-158.25$~Hz band, and $8.9 \times 10^{-25}$ for the second year in the $146.5-146.75$~Hz band. There is an overall 15\%\ calibration uncertainty on these upper limits.
No upper limits are provided in the 252 vetoed  bands, that were excluded from the coincidence analysis, since the analytical approximation would not be accurate enough. These excluded frequency bands  are marked  in  the figure.

\begin{figure}[t]
\begin{center}
\includegraphics[width=30pc]{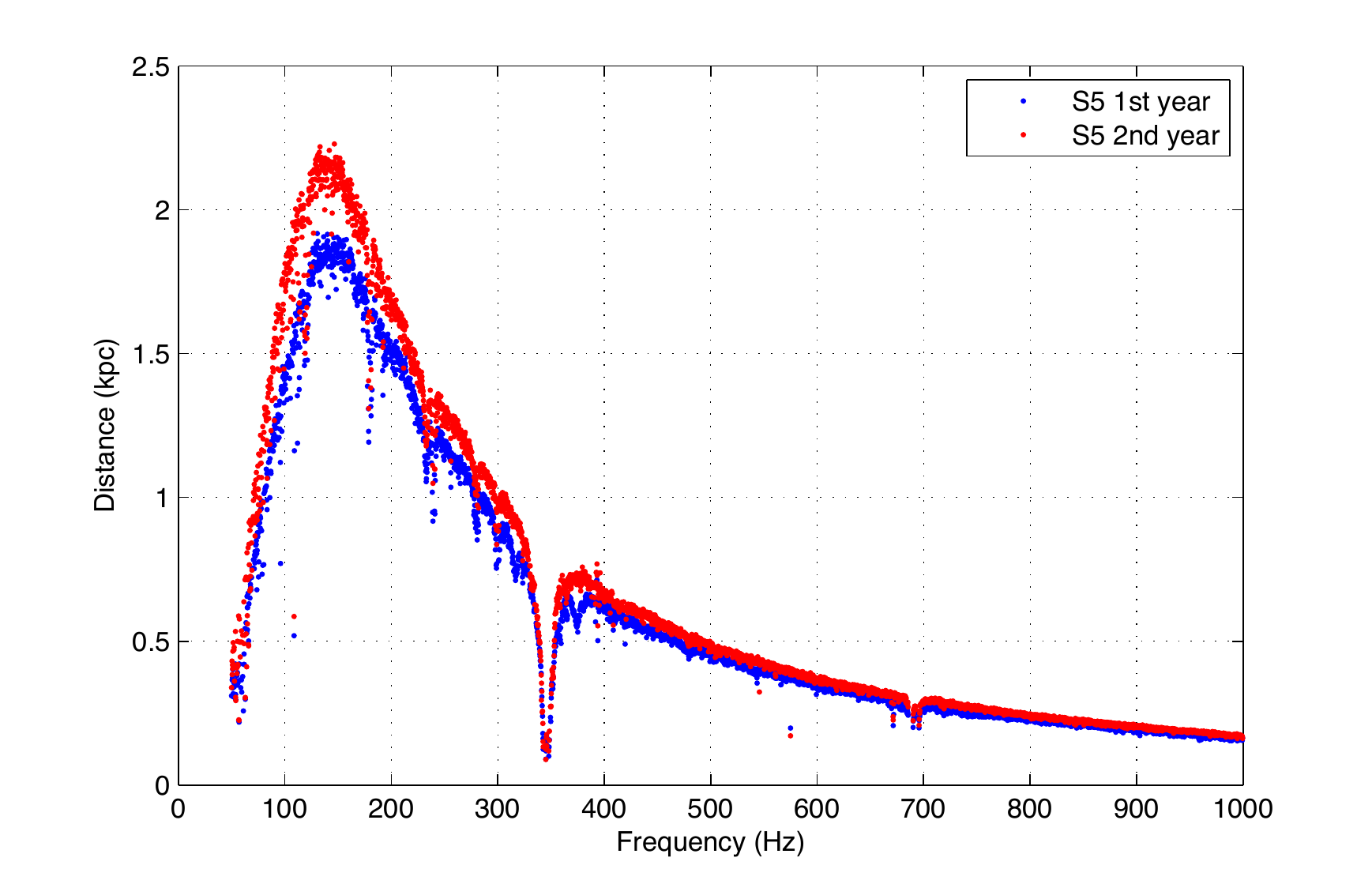}\\
\includegraphics[width=30pc]{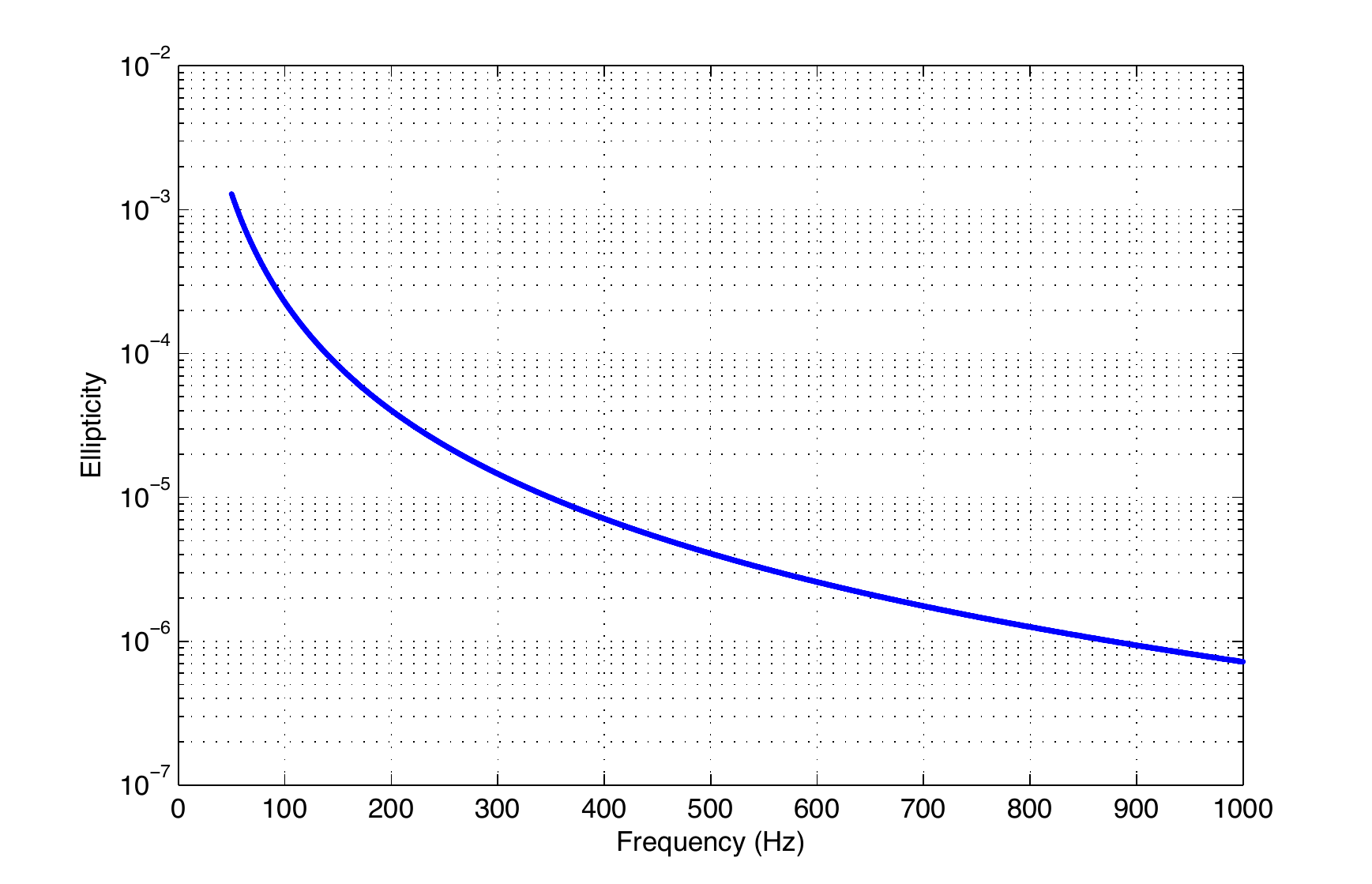}
\end{center}
\caption{\label{fig:reach} 
These plots represent the distance range (in kpc) and the maximum ellipticity, respectively, as a function of frequency. Both plots are valid for neutron stars spinning down solely due to  gravitational radiation and assuming a spin-down value of 
 $-8.9 \times 10^{-10}$~Hz/s. In the upper plot, the excluded frequency bands for which no upper limits are provided have not been considered. }
\end{figure}

Figure \ref{fig:reach} provides the maximum astrophysical reach of our search for each year of the S5 run.  The top panel shows the 
maximum distance to which we could have detected a source emitting a continuous wave signal with strain amplitude $h_0^{95\%}$.
The bottom  panel does not depend on any result from the search. It shows the corresponding ellipticity values as a function of frequency. For both plots the source is assumed to be spinning down at the maximum rate considered in the search  $-8.9 \times 10^{-10}$~Hz/s, and emitting in gravitational waves all the energy lost.
This follows formulas in paper \cite{:2007tda} and assumes the canonical value of $10^{38}$ kg m$^2$ for  
$I_{zz}$ in equation (\ref{eq:GWampl}).

Around the frequencies of greatest sensitivity, we are sensitive to objects as far away as 1.9 and 2.2 kpc for the 
first and second year of S5 and with an  ellipticity $\varepsilon$ around $10^{-4}$. 
Normal neutron stars are expected to have $\varepsilon$ less than $10^{-5}$
 \cite{JohnsonMcDaniel:2012wg, Horowitz:2009ya}. 
 Such plausible value of $\varepsilon$ could be detectable by a search like this if the object were emitting at 350 Hz and at a distance no further than 750 pc. For a source of fixed ellipticity and frequency, this search had a bit less range than the 
Einstein$@$Home  search on the same data \cite{Aasi:2012fw}.


\section{Applications of the $\chi^2$ veto and hardware-injected signals. }
\label{sec:veto2}

\begin{figure}[h]
\begin{center}
\includegraphics[width=32pc]{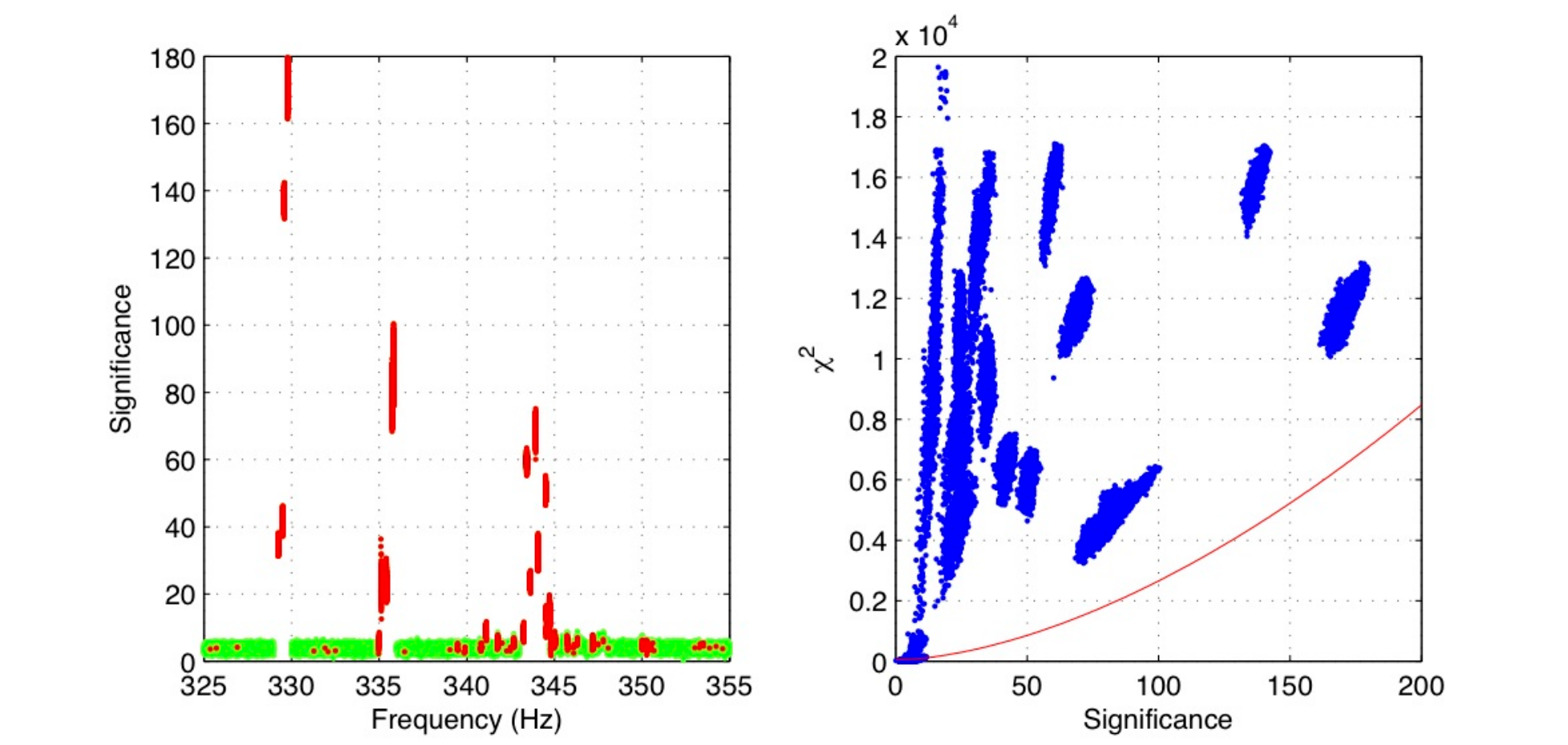} \\
\includegraphics[width=32pc]{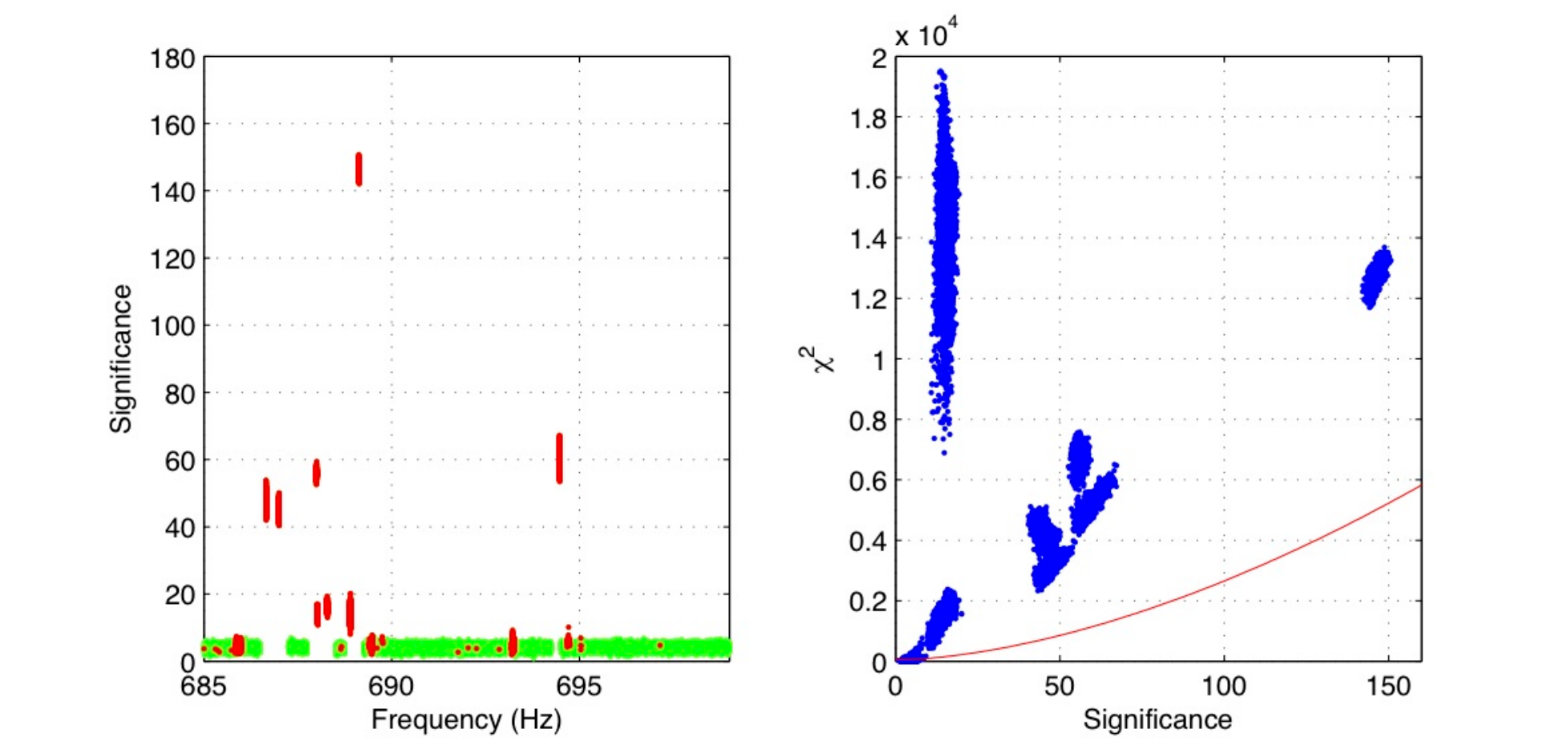} 
\end{center}
\caption{\label{fig:violins} Significance and $\chi^2$ values obtained for all the elements in the top list for the second year of S5 data for the frequency bands  325-355 Hz and  685-699 Hz.  Those two frequency bands include violin modes. Marked in dark red appear all the elements vetoed by the $\chi^2$ test. The solid line corresponds to the veto curve. }
\end{figure}

A novel feature of the search presented here is  the implementation of the  $\chi^2$ veto. It is worth mentioning that this discriminator has been able to veto all the violin modes present in the data and many other narrow instrumental artifacts. 
Figure \ref{fig:violins} demonstrates how well the  $\chi^2$ veto used works on those frequency bands affected by violin modes.

\begin{figure}[h]
\begin{center}
\includegraphics[width=32pc]{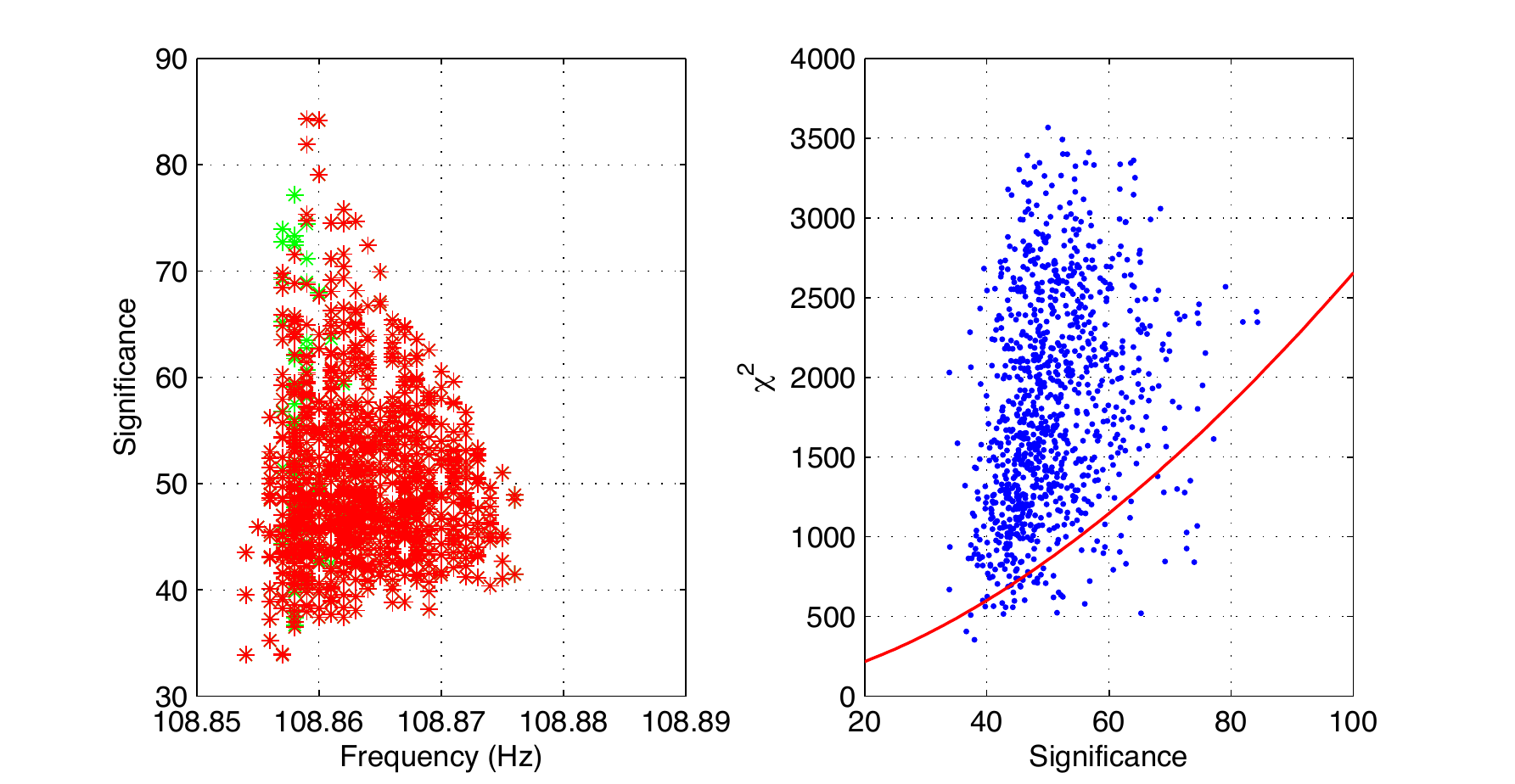} 
\end{center}
\caption{\label{fig:P3} Significance and $\chi^2$ values obtained for all the elements in the top list for the second year of S5 data for 0.25 Hz  band starting at 108.75 Hz that contains a hardware injected simulated pulsar signal. Marked in dark red appear all the elements vetoed by the $\chi^2$ test. The solid line corresponds to the veto curve. }
\end{figure}

As part of the testing and validation of search pipelines and analysis code, simulated signals are added into the interferometer length control system to produce mirror motions similar to what  would be generated if a gravitational wave signal were present. These are the so-called hardware-injected pulsars. 
During the S5 run ten artificial pulsars were injected. Four of these pulsars: P2, P3, P5 and P8,  at  frequencies  575.16,  108.85, 52.81 and  193.4 Hz respectively, were strong enough to be detected by the multi-interferometer Hough search (see  table III in \cite{Aasi:2012fw} for the detailed parameters). The hardware injections were not active all the time, having a duty factor of about $60\%$. 

The fact that these signals were not continuously present in the data caused the   $\chi^2$ test to veto most of the templates associated with them, since they did not behave like the signals we were looking for. In particular, for the second year of S5, the  elements of the top-list in frequency band containing P8 were vetoed by the $\chi^2$ test at the  $99.4\%$ level, and therefore that band was excluded from the analysis. The bands containing injected pulsars P2 and P3 were vetoed at the $87.7\%$ and $94.5\%$ level respectively, including  the most significant events. Figure \ref{fig:P3} shows the behavior of the $\chi^2$ veto for the 0.25 Hz  band starting at 108.75 Hz that contains pulsar P3. 

 In the frequency band 52.75-53.0 Hz, the candidates in the top-list were all produced by the 52.79 Hz instrumental artifact present  in H1 and consequently the search failed to detect P5. This suggests that,  in future analysis, smaller frequency intervals should be used to produce the top list of candidates, to prevent missing gravitational wave signals due to the presence of instrumental line-noise closeby.

\section{Alternative strategies and future improvements
 }
\label{sec:sensitivities}

The search presented in this paper is more robust than but not as sensitive as the hierarchical all-sky  search performed by the Einstein$@$Home  distributed computing project on the same data \cite{Aasi:2012fw}, which for example,  in the 0.5 Hz-wide band at 152.5 Hz, excluded the presence  of signals with a $h_0$  greater than  $7.6 \times 10^{-25}$ at a  90\% confidence level. This later run
used the Hough transform method to combine the information from coherent searches on a time scale of about a day
 and it was very computationally intensive. At the same time, the Einstein$@$Home search, due to its larger coherent baseline, is more sensitive to the fact that the second spin-down is not included in the search. The Hough transform method has also proven to be more robust against transient spectral disturbances than the StackSlide or PowerFlux semi-coherent methods \cite{:2007tda}.

Other strategies can be applied to perform all-sky multi-interferometer searches using the Hough transform operating on 
successive short Fourier transforms. In this section we estimate the sensitivity of the semi-coherent Hough search for two hypothetical searches to illustrate its capabilities, by either varying the duration of the total amount of data used in the multi-interferometer search, or by lowering the threshold for candidate selection.

\begin{figure}[t]
\begin{center}
\includegraphics[width=32pc]{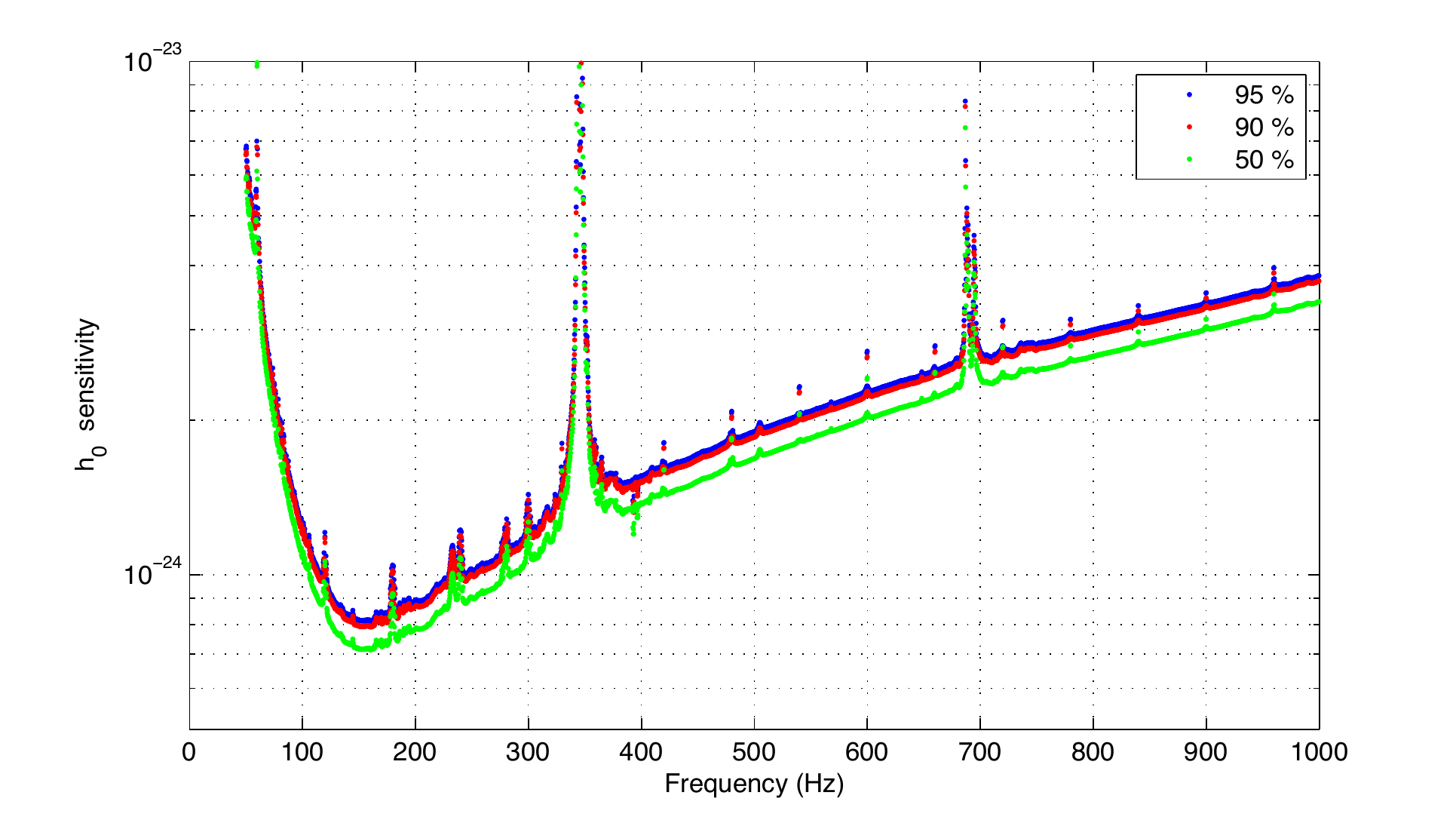} \\
\includegraphics[width=32pc]{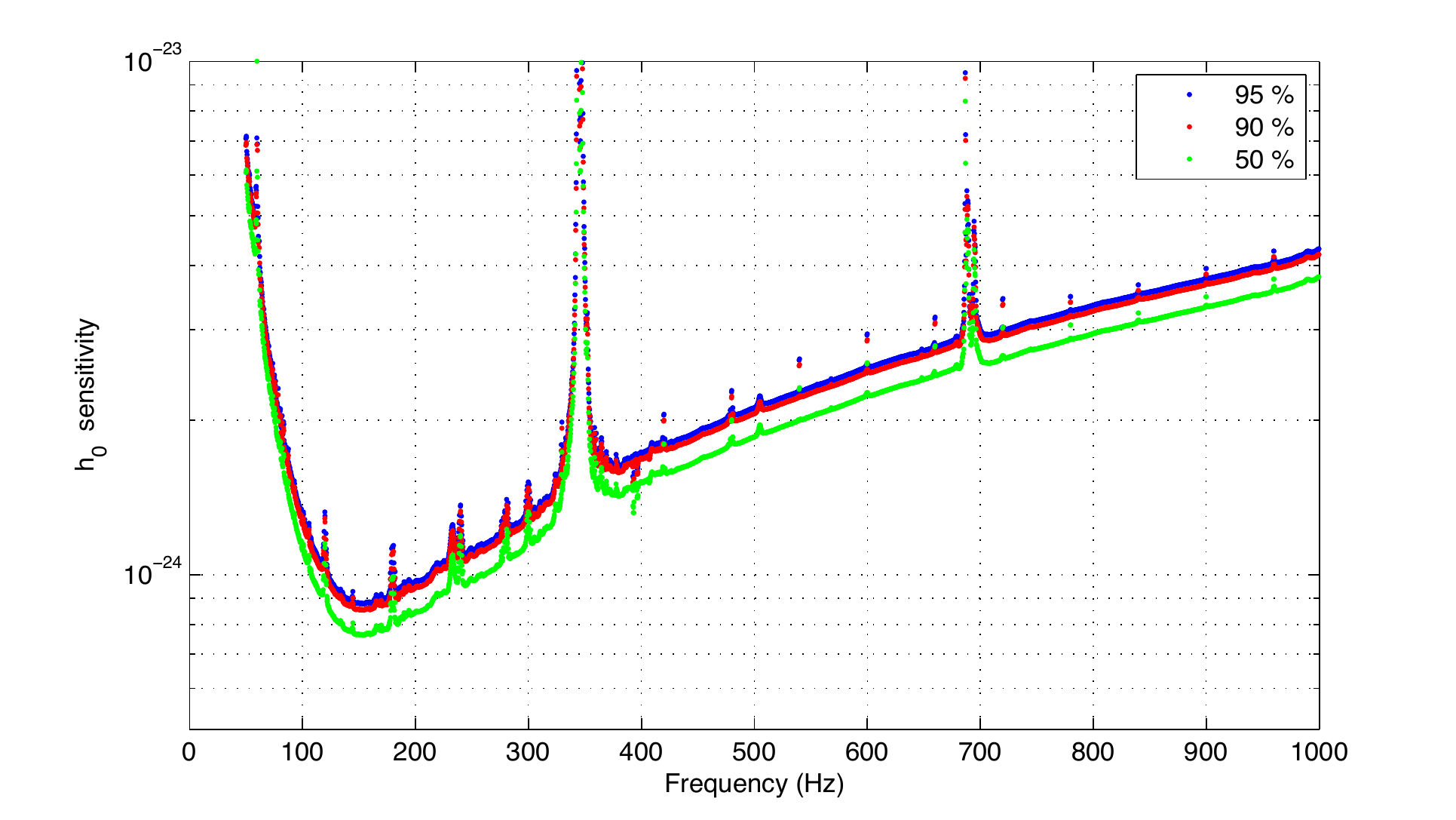} \\
\end{center}
\caption{\label{fig:sensit} Projected sensitivities at different confidence levels for (top) a combined search over the full S5 data using the same false alarm  and (bottom) sensitivity of the second year of S5 but increasing the false alarm rate.}
\end{figure}

 In the first case we consider  a search over the full S5 data with the same criteria of selecting candidates as presented here, i.e. setting a threshold in the significance   for a false alarm level equivalent to one candidate per 0.25 Hz band, but using the full data. 
This first  search would be more sensitive since we increase the number of SFTs to search over. In this case the sensitivity can be estimated from equation (\ref{eq:ul}) 
and 
 using  the desired  significance threshold as the 'loudest' event. In this case we should take into account  that the number of templates for a two years  search is double that for a single year, because of the increase of spin down values that are resolvable. This  corresponds to  the dot-dashed line in the significance threshold  in figure \ref{fig:templa}. 
Different sensitivity confidence levels can also be provided by modifying the false dismissal rate in equation (\ref{eq:S}) accordingly:
\bea
\S_{95\%}= s_\th/\sqrt{2} + \textrm{erfc}^{-1}(0.1)  \,, \nonumber\\
\S_{90\%}= s_\th/\sqrt{2} + \textrm{erfc}^{-1}(0.2)  \,, \nonumber\\
\S_{50\%}= s_\th/\sqrt{2} + \textrm{erfc}^{-1}(1)  \,.  \nonumber
\eea

In the second case, we consider only the data from  second year of S5 data but lower the threshold in the significance such that the false alarm would be 10 candidates per 0.25 Hz band (see dashed line in figure \ref{fig:templa}). 

In figure \ref{fig:sensit}  we show the projected sensitivities for these two searches for different confidence levels. The best sensitivity would be for the search performed on the combined full two years of S5 data. For example in the frequency interval 159.75-160.0 Hz the estimated sensitivity levels are of $8.1 \times 10^{-25}$, $7.9 \times 10^{-25}$  and $7.1 \times 10^{-25}$ at the $95\%$,  $90\%$, 
and $50\%$ confidence level respectively. In the second case, corresponding to an increase in false alarm rate but with a reduced amount of data, the best sensitivities are of $8.8 \times 10^{-25}$, $8.5 \times 10^{-25}$  and $7.6 \times 10^{-25}$ at the $95\%$,  $90\%$, and $50\%$ confidence level respectively. 

Notice that although both strategies are, in principle, more sensitive than the 
search presented in this paper, they would  produce many more candidates.
These would need eliminating either by demanding coincidence between 
two searches of comparable sensitivity, or by follow-up using a more sensitive, computationally intensive, search.
Coincidence analysis will be explored in future searches owing to the efficiency with which background noise is removed.
 Follow up studies will 
always be computationally limited.  Therefore the follow up capacity will actually limit 
the event threshold for candidate selection. Without the inclusion of a coincidence 
analysis, the event threshold will have to be set higher, therefore  compromising the 
potential sensitivity of the search itself.
%
%
Moreover, for the first hypothetical case, in order to achieve the projected sensitivity, one would need to perform the search over the entire  67696 SFTs available at once. If one wanted to use the ``look up table" approach to compute the Hough transform over the two years of data,  the computational cost  would increase by a factor of nine with respect to the one year search presented in this paper, and the memory usage would increase from 0.8 GB to 7.2 GB, for the same sky-patch size. The memory needs could be reduced by decreasing the sky-patch size, but at additional computational cost.
%
Another consequence of analyzing both years together is that the spin down step size in the production search would have had to be reduced significantly.



There are a number of areas where further refinements could improve the sensitivity of the Hough search. In particular, one could  decrease the grid spacing in parameter space in order to reduce the maximum mismatch allowed, increase the duration of the SFTs to increase the signal to noise ratio within a single SFT, the development of further veto strategies to increase the overall efficiency of the analysis, as well as the tracking and establishing of appropriate data-cleaning strategies to remove  narrow-band disturbances present in the peak-grams \cite{Coughlin,1742-6596-363-1-012037,1742-6596-363-1-012024}.
Several of these ideas are being addressed  and will be implemented in the ``Frequency Hough all-sky search" using data from the Virgo second and fourth science runs to analyze data between 20 and 128 Hz.

\section{Conclusions}
\label{sec:end}

In summary, we have reported the results  of an all-sky search for continuous, nearly monochromatic gravitational waves on data from LIGO's fifth science run, using a new detection pipeline based on the Hough transform. The search covered  the frequency range $50\,$--$\,1000$~Hz and with the frequency's time derivative in the range $-8.9 \times 10^{-10}$~Hz/s to zero. Since no evidence for gravitational waves has been observed,  we have derived upper limits on the intrinsic gravitational wave strain amplitude using a standard population-based method.
The best upper limits correspond to $1.0\times 10^{-24}$ for the first year of S5 in the $158-158.25$~Hz band, and $8.9 \times 10^{-25}$ for the second year in the $146.5-146.75$~Hz band (see figure \ref{fig:UL}).

This new search pipeline has allowed to process outliers down to significance from 5.10 at 50 Hz to 6.13 at 1000 Hz permitting deeper searchers than in previous Hough all-sky searches  \cite{:2007tda}.  
A set of new features have been included into the multi-detector Hough search code to be able to cope with large amounts of data
and the memory limitations on the machines. 
In addition, a   $\chi^2$ veto has been applied for the first time for continuous gravitational wave searches. This veto might be very useful for the analysis of the most recent set of data produced by the LIGO and Virgo interferometers (science runs S6, VSR2 and VSR4) whose data at lower frequencies are characterized by larger contamination of non-Gaussian noise than for S5.

 Although the search presented here is not the most sensitive one on the same  S5 data, this paper shows the potential of the new pipeline given the advantage of the lower computational cost of the Hough search and its robustness compared to other methods, and  suggests further improvements to increase the sensitivity and overall efficiency of the analysis.

\ack
\input L-V_ack_Feb2012.tex
This document  has been assigned LIGO Laboratory document number
LIGO-P1300071.


\section{References}
\bibliography{HoughS5_CQG_v1}

\end{document}

%% file: authors_LSC_Feb2013_Virgo_Jul2013_iop.tex
\author{%
J.~Aasi$^{1}$,
J.~Abadie$^{1}$,
B.~P.~Abbott$^{1}$,
R.~Abbott$^{1}$,
T.~Abbott$^{2}$,
M.~R.~Abernathy$^{1}$,
T.~Accadia$^{3}$,
F.~Acernese$^{4,5}$,
C.~Adams$^{6}$,
T.~Adams$^{7}$
% ...
}

\address {$^{1}$LIGO - California Institute of Technology, Pasadena, CA 91125, USA }
\address {$^{2}$Louisiana State University, Baton Rouge, LA 70803, USA }
\address {$^{3}$Laboratoire d'Annecy-le-Vieux de Physique des Particules (LAPP), Universit\'e de Savoie, CNRS/IN2P3, F-74941 Annecy-le-Vieux, France }
\address {$^{4}$INFN, Sezione di Napoli, Complesso Universitario di Monte S.Angelo, I-80126 Napoli, Italy }
\address {$^{5}$Universit\`a di Salerno, Fisciano, I-84084 Salerno, Italy }
\address {$^{6}$LIGO - Livingston Observatory, Livingston, LA 70754, USA }
\address {$^{7}$Cardiff University, Cardiff, CF24 3AA, United Kingdom }
% ...


%% file: L-V_ack_Feb2012.tex
The authors gratefully acknowledge the support of the United States
National Science Foundation for the construction and operation of the
LIGO Laboratory, the Science and Technology Facilities Council of the
United Kingdom, the Max-Planck-Society, and the State of
Niedersachsen/Germany for support of the construction and operation of
the GEO600 detector, and the Italian Istituto Nazionale di Fisica
Nucleare and the French Centre National de la Recherche Scientifique
for the construction and operation of the Virgo detector. The authors
also gratefully acknowledge the support of the research by these
agencies and by the Australian Research Council, 
the International Science Linkages program of the Commonwealth of Australia,
the Council of Scientific and Industrial Research of India, 
the Istituto Nazionale di Fisica Nucleare of Italy, 
the Spanish Ministerio de Econom\'ia y Competitividad,
the Conselleria d'Economia Hisenda i Innovaci\'o of the
Govern de les Illes Balears, the Foundation for Fundamental Research
on Matter supported by the Netherlands Organisation for Scientific Research, 
the Polish Ministry of Science and Higher Education, the FOCUS
Programme of Foundation for Polish Science,
the Royal Society, the Scottish Funding Council, the
Scottish Universities Physics Alliance, The National Aeronautics and
Space Administration, the Carnegie Trust, the Leverhulme Trust, the
David and Lucile Packard Foundation, the Research Corporation, and
the Alfred P. Sloan Foundation.